\documentstyle[eqsecnum,aps,epsf
]{revtex}
\newcommand{\meanr}[1]{\left\langle\,#1\,\right\rangle^{{\bf x}_0}}
\newcommand{\meanrp}[1]{\left\langle\,#1\,\right\rangle^{{\bf p}_0,{\bf x}_0}_{\bf \Omega}}
\newcommand{\meanzero}[1]{\left\langle\,#1\,\right\rangle^{\bf 0}_{\bf \Omega}}
\newcommand{\cumrp}[1]{\left\langle\,#1\,\right\rangle^{{\bf p}_0,{\bf x}_0}_{{\bf \Omega},c}}
\newcommand{\DS}{\displaystyle}
\newcommand{\tx}{\tilde{x}}
\newcommand{\ty}{\tilde{y}}
\newcommand{\tz}{\tilde{z}}
\newcommand{\tr}{\tilde{\bf x}}
\newcommand{\tpx}{\tilde{p}_x}
\newcommand{\tpy}{\tilde{p}_y}
\newcommand{\tpz}{\tilde{p}_z}
\newcommand{\tp}{\tilde{\bf p}}
\newcommand{\dg}{{^\bullet G}}
\newcommand{\gd}{{G^\bullet}}
\newcommand{\dgd}{{^\bullet G^\bullet}}
\newcommand{\tdgd}{{^\bullet \tilde{G}^\bullet}}
\newcommand{\osa}{\Omega_{\perp 1}}
\newcommand{\osb}{\Omega_{\perp 2}}
\newcommand{\os}{\Omega_{\perp}}
\newcommand{\op}{\Omega_{\parallel}}
\newcommand{\bold}[1]{\mbox{\boldmath$#1$}}
\begin{document}
\title{
Variational Approach to Hydrogen Atom in Uniform\\
Magnetic Field of Arbitrary Strength}
\author{M. Bachmann, H. Kleinert, and A. Pelster}
\address{Institut f\"ur Theoretische Physik, Freie Universit\"at Berlin,
Arnimallee 14, 14195 Berlin}
\date{\today}
\maketitle
\begin{abstract}
Extending the Feynman-Kleinert variational approach, we calculate
the temperature-dependent
effective classical potential governing the quantum statistics of a hydrogen atom 
in a uniform magnetic at all temperatures. 
The zero-temperature limit yields the binding energy of the electron which is quite accurate
for all magnetic field strengths and exhibits, in particular, the correct logarithmic
growth at large fields. 
\end{abstract}
\section{Introduction}
The quantum statistical and quantum mechanical properties of a hydrogen
atom in an external magnetic field are not exactly calculable.
Perturbative approaches yield good results only for
weak uniform fields as discussed in detail by Le Guillou and 
Zinn-Justin~\cite{zinn-justin}, who interpolated with analytic mapping techniques
the ground state energy between weak- and strong-field.
Other approaches are based on recursive procedures in higher-order perturbation theory 
\cite{cizek1,cizek2,gani}. 
Zero-temperature properties were also investigated with the help of an operator
optimization method in a second-quantized variational procedure~\cite{feranshuk}.
The behaviour at high uniform fields was inferred from treatments
of the one-dimensional hydrogen atom~\cite{landau,loudon,haines}.
Hydrogen in strong magnetic fields is still a problem under investigation, since its solution
is necessary to understand the properties of 
white dwarfs and neutron stars, as emphasized in Refs.~\cite{cohen,ivanov,heyl}. 

A compact and detailed presentation of 
the bound states and highly accurate numerically values for the
energy levels is given in Ref.~\cite{wunner}. 

Equations for a first-order
variational approach to the ground-state energy of hydrogen in a uniform magnetic field
based on the Jensen-Peierls inequality were written down a long time ago~\cite{dev0}, but
never evaluated. Apparently, they merely served as a preparation for attacking the
more complicated problem of a polaron in a magnetic field~\cite{dev0,dev1,dev2}. 

In our approach, we calculate the quantum statistical properties of the system by an
extension of variational perturbation theory~\cite{PI}.
The crucial quantity is the effective classical potential.
In the zero-temperature limit, it yields the ground state energy. Our calculations
in a magnetic field require
an extension of the formalism in Ref.~\cite{PI} which
derives the effective classical potential from
the phase space representation of the partition function.

Variational perturbation theory has an important 
advantage over other approaches: The calculation yields a good effective 
classical potential for {\it all} temperatures and coupling strengths. 
The quantum statistical partition function is obtained from a simple integral over a
Boltzmann-factor involving the effective classical potential.  
The ground
state energy is then obtained from its zero-temperature limit. The asymptotic behaviour 
in the strong-coupling limit is emerging automatically and
does not have to be derived from other sources.
\section{Effective Classical Representations for the Quantum Statistical Partition Function}
\label{effrep}
A point particle in $D$ dimensions with a potential $V({\bf x})$ and a vector 
potential ${\bf A}({\bf x})$ is described by 
a Hamiltonian
\begin{equation}
\label{ham00}
H({\bf p},{\bf x})=\frac{1}{2M}\left[{\bf p}-\frac{e}{c}{\bf A}({\bf x})\right]^2+V({\bf x}).
\end{equation}
The quantum statistical partition function is given by the euclidean
phase space path integral
\begin{equation}
\label{ham01}
Z=\oint {\cal D'}^Dx {\cal D}^Dp\,e^{-{\cal A}[{\bf p},{\bf x}]/\hbar}
\end{equation}
with an action
\begin{equation}
\label{ham02}
{\cal A}[{\bf p},{\bf x}]=\int_0^{\hbar\beta}d\tau\left[-i{\bf p}(\tau)\cdot\dot{\bf x}(\tau) 
+H({\bf p}(\tau),{\bf x}(\tau))\right],
\end{equation}
and the path measure
\begin{equation}
\label{ham03}
\oint{\cal D'}^Dx {\cal D}^Dp=\lim_{N\to\infty}\prod\limits_{n=1}^{N+1}
\left[\int\frac{d^Dx_nd^Dp_n}{(2\pi\hbar)^D} \right].
\end{equation}
The parameter $\beta=1/k_BT$ denotes the usual inverse thermal energy at 
temperature $T$, where $k_B$ is the Boltzmann constant. From $Z$ we obtain the free energy
of the system: 
\begin{equation}
\label{ham04}
F=-\frac{1}{\beta}{\rm ln}\,Z.
\end{equation}
In perturbation theory, one
treats the external
potential $V({\bf x})$ as a small quantity, and expands the partition function 
into powers of $V({\bf x})$.
Such a naive
expansion is applicable only for extremely weak couplings, and has a vanishing radius
of convergence.
Convergence is achieved by
variational perturbation theory~\cite{PI}, which
yields good approximations for all potential
strengths, as we shall see in the sequel.
\subsection{Effective Classical Potential}
All quantum-mechanical systems studied so far in variational perturbation theory were 
governed by a Hamiltonian of the standard form 
\begin{equation}
\label{ham05}
H({\bf p},{\bf x})=\frac{{\bf p}^2}{2M}+V({\bf x}).
\end{equation}
The simple quadratic dependence on the momenta makes the 
momentum integrals in the path integral (\ref{ham01}) trivial.
The remaining configuration space representation of the partition function
is used to define an effective classical potential $V_{\rm eff}({\bf x}_0)$, from which
quantum mechanical partition function is found by a classically looking integral
\begin{equation}
\label{ham06}
Z=\int \frac{d^Dx_0}{\lambda_{\rm th}^{D}}\,\exp\left[-\beta V_{\rm eff}({\bf x}_0)\right],
\end{equation}
where $\lambda_{\rm th}=\sqrt{2\pi\hbar^2\beta/M}$ is the thermal wavelength.
The Boltzmann factor plays the role of a {\em local partition function} $Z^{{\bf x}_0}$, 
which is calculated from the restricted path integral
\begin{equation}
\label{ham07}
e^{-\beta V_{\rm eff}({\bf x}_0)}\equiv
Z^{{\bf x}_0}=
\lambda_{\rm th}^{D}\oint {\cal D}^Dx\,
\delta({\bf x}_0-\overline{{\bf x}(\tau)})\,e^{-{\cal A}[{\bf x}]/\hbar},
\end{equation}
with the action 
\begin{equation}
\label{ham08}
{\cal A}[{\bf x}]=\int_0^{\hbar\beta}d\tau\,\left[\frac{M}{2}\dot{\bf x}^2(\tau)+
V({\bf x}(\tau)) 
\right],
\end{equation}
and the path measure
\begin{equation}
\label{ham09}
\oint{\cal D}^Dx=\lim_{N\to\infty}\prod\limits_{n=1}^{N+1}
\left\{\int\frac{d^Dx_n}{[2\pi\hbar^2\beta/M(N+1)]^{D/2}} \right\}.
\end{equation}
The special treatment of the temporal average of the Fourier path 
\begin{equation}
\label{ham10}
{\bf x}_0=\overline{{\bf x}(\tau)}=\frac{1}{\hbar\beta}\int_0^{\hbar\beta}d\tau\,{\bf x}(\tau)
\end{equation}
is essential for the quality of the results. It
subtracts from the harmonic fluctuation width 
$\langle{\bf x}^2\rangle^{\rm cl}$ the classical 
divergence proportional to 
$T=1/k_B\beta$ of the Dulong-Petit law~\cite{PI,density}. 
Such diverging
fluctuations cannot be treated perturbatively, and require the final integration 
in expression (\ref{ham06}) to be done 
numerically.

For the
Coulomb potential $V({\bf x})=-e^2/4\pi\varepsilon_0\,|{\bf x}|$ in three dimensions, 
the effective classical potential in 
Eq.~(\ref{ham07}) can be approximated well by variational perturbation 
theory~\cite{PI,density,hatom}.
\subsection{Effective Classical Hamiltonian}
In order to deal with Hamiltonians like (\ref{ham00}) which contain a 
${\bf p}\cdot{\bf A}({\bf x})$-term,
we must generalize the variational procedure. Extending (\ref{ham07}), we define 
an {\it effective classical Hamiltonian} by the phase space path integral
\begin{equation}
\label{ham12}
e^{-\beta H_{\rm eff}({\bf p}_0,{\bf x}_0)}\equiv
Z^{{\bf p}_0,{\bf x}_0}=
(2\pi\hbar)^D\oint {\cal D'}^Dx{\cal D}^Dp\,\delta({\bf x}_0-\overline{{\bf x}(\tau)})
\delta({\bf p}_0-\overline{{\bf p}(\tau)})\,e^{-{\cal A}[{\bf p},{\bf x}]/\hbar},
\end{equation}
with the action (\ref{ham02}) and the measure~(\ref{ham03}). This allows us to express
the partition function as the classically looking phase space integral 
\begin{equation}
\label{ham11}
Z=\int\frac{d^Dx_0d^Dp_0}{(2\pi\hbar)^D}\,\exp\left[-\beta H_{\rm eff}({\bf p}_0,{\bf x}_0)
\right],
\end{equation}
where ${\bf p}_0$ is the temporal average of the momentum:
\begin{equation}
\label{ham13}
{\bf p}_0=\overline{{\bf p}(\tau)}=\frac{1}{\hbar\beta}\int_0^{\hbar\beta}d\tau\,{\bf p}(\tau).
\end{equation}
The fixing of ${\bf p}_0$ is done for the same 
reason as that for ${\bf x}_0$, 
since the classical expectation value $\langle {\bf p}^2\rangle^{\rm cl}$ 
is diverging linearly with $T$, just as $\langle{\bf x}^2 \rangle^{\rm cl}$.

In the special case of a standard 
Hamiltonian (\ref{ham05}), the effective Hamiltonian in Eq.~(\ref{ham11}) 
reduces to the effective classical potential, since
the momentum integral in 
Eq.~(\ref{ham12}) can then be easily performed, and the resulting restricted 
partition function becomes
\begin{equation}
\label{ham14}
Z^{{\bf p}_0,{\bf x}_0}=\exp\left(-\beta\frac{{\bf p}_0^2}{2M} \right)\,
Z^{{\bf x}_0}
\end{equation}  
with the local partition function $Z^{{\bf x}_0}=\exp[-\beta V_{\rm eff}({\bf x}_0)]$ 
of Eq.~(\ref{ham07}).
Thus the complete quantum statistical
partition function is given by (\ref{ham11}), with an effective classical Hamilton
function
\begin{equation}
\label{ham15}
H_{\rm eff}({\bf p}_0,{\bf x}_0)=\frac{{\bf p}_0^2}{2M}+V_{\rm eff}({\bf x}_0).
\end{equation}
As a consequence of the purely quadratic momentum dependence of $H({\bf p},{\bf x})$
in (\ref{ham05}), the 
${\bf p}_0$-integral in (\ref{ham11}) can be done, thus expressing 
the quantum statistical partition function as a pure 
configuration space integral over the Boltzmann factor involving
the effective classical potential $V_{\rm eff}({\bf x}_0)$, as in Eq.~(\ref{ham06}).
\subsection{Exact Effective Classical Hamiltonian for an Electron in a Constant Magnetic Field}
\label{exham}
The effective classical Hamiltonian for the electron moving
in a constant magnetic field can be calculated exactly. 
We consider a
magnetic field ${\bf B}=B {\bf e}_z$ pointing along the positive $z$-axis. The only nontrivial
motion of the electron is in the $x\!\!-\!\!y$-plane. 
In symmetric gauge the vector potential is given by
\begin{equation}
\label{ham16}
{\bf A}({\bf x})=\frac{B}{2}(-y,x,0).
\end{equation}
The choice of the gauge does not affect the partition function since the periodic path 
integral (\ref{ham01}) is gauge invariant.
Ignoring the trivial free particle motion along the $z$-direction, we may restrict our attention
to the two-dimensional Hamiltonian
\begin{equation}
\label{ham17}
H({\bf p},{\bf x})=\frac{{\bf p}^2}{2M}-\frac{1}{2}\omega_c l_z({\bf p},{\bf x})+
\frac{1}{8}M\omega_c^2{\bf x}^2
\end{equation}
with ${\bf x}=(x,y)$ and ${\bf p}=(p_x,p_y)$. 
Here, $\omega_c=eB/Mc$ is the Landau frequency, and 
\begin{equation}
\label{ham18}
l_z({\bf p},{\bf x})=({\bf x}\times{\bf p})_z=xp_y-yp_x
\end{equation}
the third component of the orbital angular momentum.
The partition function of the problem is given by Eq.~(\ref{ham11}), with $D=2$. 
Being interested in an effective classical formulation, we have to calculate the
path integral (\ref{ham12}). First we express the $\delta$-function for the averaged
momentum as a Fourier integral 
\begin{eqnarray}
\label{ham19}
\delta({\bf p}_0-\overline{{\bf p}(\tau)})&=&
\int \frac{d^2\xi}{(2\pi\hbar)^2}\,\exp\left(-\frac{i}{\hbar}\bold{\xi}\cdot{\bf p}_0
\right)\exp\left[-\frac{1}{\hbar}\int_0^{\hbar\beta}d\tau\,
{\bf v}_0(\bold{\xi})\cdot {\bf p}(\tau)\right]
\end{eqnarray}
involving an auxiliary source
\begin{equation}
\label{ham20}
{\bf v}_0(\bold{\xi})=-\frac{i}{\hbar\beta}\,\bold{\xi}
\end{equation}
which is constant in time. Substituting the $\delta$-function in Eq.~(\ref{ham12}) 
by this source representation, the partition function reads
\begin{eqnarray}
\label{ham21}
Z^{{\bf p}_0,{\bf x}_0}&=&\int d^2\xi\,\exp\left(-\frac{i}{\hbar}\bold{\xi}\cdot{\bf p}_0\right)
\oint {\cal D'}^2x{\cal D}^2p\,\delta({\bf x}_0-\overline{{\bf x}(\tau)})\nonumber\\
&&\times\exp\left\{-\frac{1}{\hbar}\int_0^{\hbar\beta}d\tau\,
\left[-i{\bf p}(\tau)\cdot\dot{\bf x}(\tau)+H({\bf p}(\tau),{\bf x}(\tau))
+{\bf v}_0(\bold{\xi})\cdot{\bf p}(\tau)\right]\right\}.
\end{eqnarray}
Evaluating the momentum integrals 
and utilizing the periodicity property ${\bf x}(0)={\bf x}(\hbar\beta)$, we obtain the 
configuration space path integral
\begin{eqnarray}
\label{ham22}
Z^{{\bf p}_0,{\bf x}_0}=\lim_{\Omega\to 0}
&&\int d^2\xi\,\exp\left(-\frac{i}{\hbar}\bold{\xi}\cdot{\bf p}_0-\frac{M}{2\hbar^2\beta}
\bold{\xi}^2\right)
\oint {\cal D}^2x\,\delta({\bf x}_0-\overline{{\bf x}(\tau)})\nonumber\\
&&\times\exp\left\{-\frac{1}{\hbar}\int_0^{\hbar\beta}d\tau\,\left[\frac{M}{2}\dot{\bf x}^2(\tau) 
+\frac{1}{2}M\Omega^2{\bf x}^2(\tau)
-\frac{i}{2}M\omega_c({\bf x}(\tau)\times \dot{\bf x}(\tau))_z+
{\bf x}(\tau)\cdot {\bf j}_1
(\bold{\xi})\right]\right\},
\end{eqnarray}
where the source ${\bf v}_0$ coupled to the momentum in 
(\ref{ham21}) has turned to a source ${\bf j}_1$
coupled to the path in configuration space~\cite{correlation}, with components
\begin{equation}
\label{ham23}
{\bf j}_1(\bold{\xi})=\frac{M}{2}\omega_c\left(\,v_{0y}(\bold{\xi}),-v_{0x}(\bold{\xi}) 
\,\right)=\frac{i\omega_cM}{2\hbar\beta}\left(\,-\xi_y,\xi_x \,\right).
\end{equation}
We have introduced an additional harmonic oscillator in Eq.~(\ref{ham22}) which will 
turn out to be useful at intermediate stages of the development. At the end of the calculation,
only the limit $\Omega\to 0$ will be relevant.

Expressing the $\delta$-function in the path integral of Eq.~(\ref{ham22}) by the Fourier
integral
\begin{equation}
\label{ham23b}
\delta({\bf x}_0-\overline{{\bf x}(\tau)})=\int \frac{d^2\kappa}{(2\pi)^2}\,
\exp\left(\frac{i}{\hbar}\bold{\kappa}\cdot {\bf x}_0 \right)
\exp\left[-\frac{1}{\hbar}\int_0^{\hbar\beta}d\tau\,{\bf j}_2(\bold{\kappa})\cdot
{\bf x}(\tau) \right]
\end{equation}
with the new source
\begin{equation}
\label{ham23c}
{\bf j}_2(\bold{\kappa})=\frac{i\bold{\kappa}}{\beta},
\end{equation}
the partition function (\ref{ham22}) can be written as
\begin{equation}
\label{ham23d}
Z^{{\bf p}_0,{\bf x}_0}=\lim_{\Omega\to 0}
\int d^2\xi\,\exp\left(-\frac{i}{\hbar}\bold{\xi}\cdot{\bf p}_0-\frac{M}{2\hbar^2\beta}
\bold{\xi}^2\right)\int\frac{d^2\kappa}{(2\pi)^2}\,\exp\left(\frac{i}{\hbar}
\bold{\kappa}\cdot {\bf x}_0 \right)\,Z_\Omega[{\bf J}(\bold{\xi},\bold{\kappa})].
\end{equation}
The functional $Z_\Omega[{\bf J}(\bold{\xi},\bold{\kappa})]$ is defined as 
the configuration space path integral
\begin{equation}
\label{ham23e}
Z_\Omega[{\bf J}(\bold{\xi},\bold{\kappa})]=\oint{\cal D}^2x\,
\exp\left[-\frac{1}{2}\int_0^{\hbar\beta}d\tau\int_0^{\hbar\beta}d\tau'\,
{\bf x}(\tau)\,{\bf G}^{-1}(\tau,\tau')\,{\bf x}(\tau')-\frac{1}{\hbar}\int_0^{\hbar\beta}
d\tau\,{\bf J}(\bold{\xi},\bold{\kappa})\cdot{\bf x}(\tau) \right],
\end{equation}
where we have introduced the combined source 
${\bf J}(\bold{\xi},\bold{\kappa})={\bf j}_1(\bold{\xi})+{\bf j}_2(\bold{\kappa})$.
Formally, the solution reads
\begin{equation}
\label{ham23ee}
Z_\Omega[{\bf J}(\bold{\xi},\bold{\kappa})]=Z_\Omega[0]
\exp\left[-\frac{1}{2\hbar^2}\int_0^{\hbar\beta}d\tau\int_0^{\hbar\beta}d\tau'\,
{\bf J}(\bold{\xi},\bold{\kappa})\,{\bf G}(\tau,\tau')\,{\bf J}(\bold{\xi},\bold{\kappa}) \right],
\end{equation}
where ${\bf G}(\tau,\tau')$ is the matrix of Green functions obtained by inverting
\begin{equation}
\label{ham23f}
{\bf G}^{-1}(\tau,\tau')=\frac{M}{\hbar}\,\left(\begin{array}{cc}
-\frac{d^2}{d \tau^2}+\Omega^2 & -i\omega_c\frac{d}{d\tau}\\
i\omega_c\frac{d}{d\tau} & -\frac{d^2}{d \tau^2}+\Omega^2
\end{array}\right)\,\delta(\tau-\tau').
\end{equation}
The inversion is easily done in frequency space after spectrally 
decomposing the $\delta$-function into 
the Matsubara frequencies $\omega_m=2\pi m/\hbar\beta$,
\begin{equation}
\label{ham23g}
\delta(\tau-\tau')=\frac{1}{\hbar\beta}\sum\limits_{m=-\infty}^\infty\,e^{i\omega_m(\tau-\tau')}.
\end{equation}
The result is
\begin{equation}
\label{ham23h}
\tilde{\bf G}(\omega_m)=\frac{\hbar}{M}\frac{1}{{\rm det}\,\tilde{\bf G}}
\left(\begin{array}{cc}\omega_m^2+\Omega^2 & -\omega_c\omega_m\\
 \omega_c\omega_m & \omega_m^2+\Omega^2\end{array}\right).
\end{equation}
At this point, the additional
oscillator in Eq.~(\ref{ham22}) proves useful: It ensures that the
determinant
\begin{equation}
\label{ham23i}
{\rm det}\,\tilde{\bf G}(\omega_m)=(\omega_m^2+\Omega^2)^2+\omega_c^2\omega_m^2
\end{equation}
is nonzero for $m=0$, thus playing the role of an infrared regulator.
The Fourier expansion
\begin{equation}
\label{ham23j}
{\bf G}(\tau,\tau')=\frac{1}{\hbar\beta}\sum\limits_{m=-\infty}^\infty\,\tilde{\bf G}(\omega_m)
e^{-i\omega_m(\tau-\tau')}
\end{equation}
yields the matrix of Green functions
\begin{equation}
\label{ham23k}
{\bf G}(\tau,\tau')=\left(\begin{array}{cc}G_{xx}(\tau,\tau') & G_{xy}(\tau,\tau')\\
G_{yx}(\tau,\tau') & G_{yy}(\tau,\tau')  \end{array}\right)
\end{equation}
which inherits the symmetry properties from the kernel (\ref{ham23f}):
\begin{equation}
\label{ham23l}
G_{xx}(\tau,\tau')=G_{yy}(\tau,\tau'),\qquad G_{xy}(\tau,\tau')=
-G_{yx}(\tau,\tau').
\end{equation} 
A more detailed description of these Green functions is given in 
Apps.~\ref{appgen} and \ref{appprop}.

Since the current ${\bf J}$ does not depend on the euclidean time, the expression
(\ref{ham23ee}) simplifies therefore to
\begin{equation}
\label{ham23m}
Z_\Omega[{\bf J}(\bold{\xi},\bold{\kappa})]=Z_\Omega[0]
\exp\left[-\frac{1}{\hbar^2}{\bf J}^2(\bold{\xi},\bold{\kappa})
\int_0^{\hbar\beta}d\tau\int_0^{\hbar\beta}d\tau'\,G_{xx}(\tau,\tau')\right].
\end{equation}
The Green function has the Fourier decomposition
\begin{equation}
\label{ham23n}
G_{xx}(\tau,\tau')=\frac{1}{M\beta}\sum\limits_{m=-\infty}^\infty\,
\frac{\omega_m^2+\Omega^2}{(\omega_m^2+\Omega_+^2)(\omega_m^2+\Omega_-^2)}
\,e^{-i\omega_m(\tau-\tau')},
\end{equation}
where $\Omega_\pm$ are the frequencies
\begin{equation}
\label{ham23o}
\Omega_\pm=\sqrt{\Omega^2+\frac{1}{2}\omega_c^2\pm\omega_c\sqrt{\Omega^2+\frac{1}{4}\omega_c^2}}.
\end{equation}
The ratios in the sum of (\ref{ham23n}) can be decomposed into two partial 
fractions, each of them 
representing a single harmonic oscillator with frequency $\Omega_+$ and $\Omega_-$, respectively.
The analytic form of the periodic Green function of a single harmonic oscillator is well 
known (see Chap. 3 in \cite{PI}), and we obtain for the present Green function (\ref{ham23o}):
\begin{equation}
\label{ham23p}
G_{xx}(\tau,\tau')=\frac{1}{M\beta}\left(\frac{\hbar\beta}{2\Omega_+}\frac{\Omega_+^2-\Omega^2}
{\Omega_+^2-\Omega_-^2}\frac{\cosh{\Omega_+(|\tau-\tau'|-\hbar\beta/2)}}
{\sinh{\hbar\beta\Omega_+/2}}- \frac{\hbar\beta}{2\Omega_-}\frac{\Omega_-^2-\Omega^2}
{\Omega_+^2-\Omega_-^2}\frac{\cosh{\Omega_-(|\tau-\tau'|-\hbar\beta/2)}}
{\sinh{\hbar\beta\Omega_-/2}}\right).
\end{equation}
By writing the determinant (\ref{ham23i}) as
\begin{equation}
\label{ham23ii}
{\rm det}\,\tilde{\bf G}(\omega_m)=(\omega_m^2+\Omega_+^2)(\omega_m^2+\Omega_-^2)
\end{equation}
and summing over the logarithms of this, we calculate
the partition function as a product of two single harmonic
oscillators:
\begin{equation}
\label{ham23q}
Z_\Omega=Z_\Omega[0]=
\frac{1}{2\sinh{\hbar\beta\Omega_+/2}}\frac{1}{2\sinh{\hbar\beta\Omega_-/2}}.
\end{equation}
The results (\ref{ham23p}) and (\ref{ham23q}) determine the generating functional (\ref{ham23m}).
The euclidean time integrations are then easily done, and subsequently the $\bold{\kappa}$- and
$\bold{\xi}$-integrations in (\ref{ham23d}). As a result, we obtain the restricted 
partition function
\begin{equation}
\label{ham25}
Z^{{\bf p}_0,{\bf x}_0}=\lim_{\Omega\to 0}\exp\left\{-\beta\left(-\frac{1}{\beta}{\rm ln}\,
\frac{\sinh \hbar\beta\Omega_+/2}{\hbar\beta\Omega_+/2}
\frac{\sinh \hbar\beta\Omega_-/2}{\hbar\beta\Omega_-/2}
 +\frac{{\bf p}_0^2}{2M}-\frac{1}{2}\omega_c l_z({\bf p}_0,{\bf x}_0)
+\frac{1}{8}M\omega_c^2{\bf x}_0^2+\frac{1}{2}M\Omega^2{\bf x}_0^2
\right)\right\}.
\end{equation}
If we now remove the additional oscillator by
taking the limit $\Omega\to 0$, we find from (\ref{ham23o}): $\Omega_+\rightarrow \omega_c$,
$\Omega_-\rightarrow 0$, and therefore
\begin{equation}
\label{ham26}
\lim_{\Omega\to 0}\frac{\sinh{\hbar\beta\Omega_+/2}}{\hbar\beta\Omega_+/2}=
\frac{\sinh{\hbar\beta\omega_c/2}}{\hbar\beta\omega_c/2},\quad 
\lim_{\Omega\to 0}\frac{\sinh{\hbar\beta\Omega_-/2}}{\hbar\beta\Omega_-/2}=1.
\end{equation}
Recalling the definition (\ref{ham12}), we identify
the exact effective classical Hamiltonian for an electron in a magnetic field as
\begin{equation}
\label{ham27}
H_{\rm eff}({\bf p}_0,{\bf x}_0)=\frac{1}{\beta}\,{\rm ln}\,\frac{\sinh{\hbar\beta\omega_c/2}}
{\hbar\beta\omega_c/2}+\frac{{\bf p}_0^2}{2M}-\frac{1}{2}\omega_c\,l_z({\bf p}_0,{\bf x}_0)
+\frac{1}{8}M\omega_c^2{\bf x}_0^2.
\end{equation}
Integrating out the momenta in Eq.~(\ref{ham11}),
the configuration space representation (\ref{ham06}) for the partition function 
contains the effective classical potential for a charged particle in the plane
perpendicular to the direction of a uniform magnetic field
\begin{equation}
\label{ham28}
V_{\rm eff}({\bf x}_0)=\frac{1}{\beta}\,{\rm ln}\,\frac{\sinh{\hbar\beta\omega_c/2}}
{\hbar\beta\omega_c/2}.
\end{equation}
Note that this is a constant potential.

Denoting the area $\int d^2x_0$ by $A$, we find the exact quantum statistical partition function
\begin{equation}
\label{ham29}
Z=\frac{A}{\lambda_{\rm th}^2}\frac{\hbar\beta\omega_c/2}{\sinh{\hbar\beta\omega_c/2}}.
\end{equation}
After these preparations, we can turn our attention to the system we want to study in this paper:
the hydrogen atom in a uniform magnetic field,
where the additional Coulomb interaction prevents us from finding 
an exact solution for the effective classical Hamilton function.
\section{Hydrogen Atom in Constant Magnetic Field}
The zero-temperature properties of the hydrogen atom without external fiels are exactly known. 
For the quantum statistics at finite temperatures, an analytic expression exists, but it is hard
to evaluate. It is easier to find an accurate 
approximative result with the help of variational perturbation
theory~\cite{hatom}. Similar calculations
have been performed for the electron-proton pair distribution function which
can be interpreted as the unnormalized density matrix~\cite{density}. 

Here we extend this method of calculation to the hydrogen atom in a constant magnetic field.
This extension is quite nontrivial
since the weak- and strong-field limits will turn out to exhibit completely different
asymptotic behaviours. 
Let us first generalize
variational perturbation theory to an electron in a constant magnetic field and arbitrary
potential. 
\subsection{Generalized Variational Perturbation Theory}
We consider once more the effective classical form (\ref{ham11}) of the quantum statistical 
partition function, which requires the path integration (\ref{ham12}) in phase space. 
Fluctuations parallel and vertical to the magnetic field lines are now both nontrivial, 
and we must deal with the full 
three-dimensional system and the components of the electron position and momentum 
are now denoted by ${\bf x}=(x,y,z)$ and ${\bf p}=(p_x,p_y,p_z)$.
For the uniform magnetic field pointing
along the $z$-axis, the vector potential ${\bf A}({\bf x})$ is used in the gauge (\ref{ham16}).  
Thus the Hamilton function of an electron in a magnetic field and an 
arbitrary potential $V({\bf x})$ is
\begin{equation}
\label{vpt00}
H({\bf p},{\bf x})=\frac{{\bf p}^2}{2M}-\frac{1}{2}\omega_cl_z({\bf p},{\bf x})+\frac{1}{8}
M\omega_c^2{\bf x}^2+V({\bf x}).
\end{equation}
 The orbital angular momentum 
$l_z({\bf p},{\bf x})$ was introduced in Eq.~(\ref{ham18}), and the Landau frequency
$\omega_c$ below Eq.~(\ref{ham17}). The importance of the separation of the zero 
frequency components ${\bf x}_0$ and ${\bf p}_0$ was discussed 
in Sect.~\ref{effrep}. Their divergence with the temperature $T$ prevents a perturbative
treatment. Thus it is essential to set up the perturbation theory only for 
the fluctuations around ${\bf x}_0$ and ${\bf p}_0$. For this we rewrite the action functional
(\ref{ham02}) associated with the Hamiltonian (\ref{vpt00}) as
\begin{equation}
\label{vpt01}
{\cal A}[{\bf p},{\bf x}]=
{\cal A}^{{\bf p}_0,{\bf x}_0}_{\bf \Omega}[{\bf p},{\bf x}]
+{\cal A}_{\rm int}[{\bf p},{\bf x}],
\end{equation}
where we have introduced the fluctuation action 
\begin{eqnarray}
\label{vpt02}
{\cal A}^{{\bf p}_0,{\bf x}_0}_{\bf \Omega}[{\bf p},{\bf x}]=
\int_0^{\hbar\beta}d\tau\, \Big\{&&-i[{\bf p}(\tau)-{\bf p}_0]\cdot\dot{\bf x}(\tau)+
\frac{1}{2M}[{\bf p}(\tau)-{\bf p}_0]^2+\frac{1}{2}\osa 
l_z({\bf p}(\tau)-{\bf p}_0,{\bf x}(\tau)-{\bf x}_0)\nonumber\\
&&+\frac{1}{8}M\osb^2
\left[{\bf x}^\perp(\tau)-{\bf x}_0^\perp\right]^2+\frac{1}{2}M\op^2[z(\tau)-z_0]^2 \Big\},
\end{eqnarray}
in which ${\bf x}^\perp=(x,y)$ denotes the transverse part of ${\bf x}$.
The interaction is now
\begin{equation}
\label{vpt03}
{\cal A}_{\rm int}[{\bf p},{\bf x}]=\int_0^{\hbar\beta}d\tau\,
V_{\rm int}({\bf p}(\tau),{\bf x}(\tau))
={\cal A}[{\bf p},{\bf x}]-
{\cal A}^{{\bf p}_0,{\bf x}_0}_{\bf \Omega}[{\bf p},{\bf x}]
\end{equation}
with the interaction potential 
\begin{eqnarray}
\label{vpt03b}
V_{\rm int}({\bf p}(\tau),{\bf x}(\tau))&=&\frac{1}{2M}\left\{{\bf p}^2(\tau)-
\left[{\bf p}(\tau)-{\bf p}_0 \right]^2 \right\}+\frac{1}{2}\omega_c\,
{\bf p}^\perp(\tau)\times
{\bf x}^\perp(\tau)
\nonumber\\
&&-\frac{1}{2}\osa
({\bf p}^\perp(\tau)-{\bf p}^\perp_0)\times
({\bf x}^\perp(\tau)-{\bf x}^\perp_0)
+\frac{1}{8}M\omega_c^2{{\bf x}^\perp}^2(\tau)\nonumber\\
&&-\frac{1}{8}M\osb^2
\left[{\bf x}^\perp(\tau)-{\bf x}_0^\perp\right]^2
-\frac{1}{2}M\op^2[z(\tau)-z_0]^2+V({\bf x}(\tau)),
\end{eqnarray} 
where ${\bf p}^\perp=(p_x,p_y)$.
The frequencies 
${\bf \Omega}=(\osa,\osb,\op)$ are for the moment arbitrary.
The decomposition (\ref{vpt01}) forms the basis for the variational approach, where
the first term in the action~(\ref{vpt01}) allows an 
exact treatment. The transverse part was given in Sec.~\ref{exham} and the longitudinal
part is trivial, since it is harmonic with frequency $\op$. 
The associated partition function is given by the path integral 
\begin{equation}
\label{vpt04}
Z^{{\bf p}_0,{\bf x}_0}_{\bf \Omega}=
\oint{\cal D'}^3x{\cal D}^3p\,\delta({\bf x}_0-\overline{{\bf x}(\tau)})
\delta({\bf p}_0-\overline{{\bf p}(\tau)})
e^{-{\cal A}^{{\bf p}_0,{\bf x}_0}_{\bf \Omega}[{\bf p},{\bf x}]/\hbar},
\end{equation}
which can be performed. Details are given in Appendix~\ref{appgenrp}. The result is
\begin{equation}
\label{vpt04b}
Z^{{\bf p}_0,{\bf x}_0}_{\bf \Omega}=
\frac{\hbar\beta\Omega_+/2}{\sinh{\hbar\beta\Omega_+/2}}\,
\frac{\hbar\beta\Omega_-/2}{\sinh{\hbar\beta\Omega_-/2}}\,
\frac{\hbar\beta\op/2}{\sinh{\hbar\beta\op/2}},
\end{equation}
where auxiliary frequencies are composed of the frequencies $\osa,\osb$ in the action 
(\ref{vpt02}) as
\begin{equation}
\label{vpt04c}
\Omega_{\pm}(\osa,\osb)=\frac{1}{2}\,|\osa\pm\osb|.
\end{equation}
This partition function serves in the subsequent
pertubation expansion as trial system which depends explicitly on the frequencies 
${\bf \Omega}$. The correlation functions are a straightforward generalization
of (\ref{ham23k}) to three dimensions:
\begin{equation}
\label{hyd05c}
{\bf G}^{{\bf x}_0}(\tau,\tau')=\left(\begin{array}{ccc}
G_{xx}^{{\bf x}_0}(\tau,\tau') & G_{xy}^{{\bf x}_0}(\tau,\tau') & 0\\
G_{yx}^{{\bf x}_0}(\tau,\tau') & G_{yy}^{{\bf x}_0}(\tau,\tau') & 0\\
0 & 0 & G_{zz}^{{\bf x}_0}(\tau,\tau')\end{array}\right),
\end{equation}
whose explicit form is derived in App.~\ref{appgenrp}.

The ${\bf \Omega}$-dependent action in Eq.~(\ref{vpt01}) is treated perturbatively.
Writing the partition function (\ref{ham12}) as
\begin{equation}
\label{vpt05}
Z^{{\bf p}_0,{\bf x}_0}=
(2\pi\hbar)^3\oint {\cal D'}^3x{\cal D}^3p\,\delta({\bf x}_0-\overline{{\bf x}(\tau)})
\delta({\bf p}_0-\overline{{\bf p}(\tau)})\,
\exp\left\{-\frac{1}{\hbar}{\cal A}^{{\bf p}_0,{\bf x}_0}_{\bf \Omega}
[{\bf p},{\bf x}]\right\}
\exp\left\{-\frac{1}{\hbar}\int_0^{\hbar\beta}d\tau\,
V_{\rm int}({\bf p}(\tau),{\bf x}(\tau))\right\},
\end{equation}
the second exponential is expanded into a Taylor series, yielding
\begin{eqnarray}
\label{vpt06}
Z^{{\bf p}_0,{\bf x}_0}&=&
(2\pi\hbar)^3\oint {\cal D'}^3x{\cal D}^3p\,\delta({\bf x}_0-\overline{{\bf x}(\tau)})
\delta({\bf p}_0-\overline{{\bf p}(\tau)})\,
\exp\left\{-\frac{1}{\hbar}{\cal A}^{{\bf p}_0,{\bf x}_0}_{\bf \Omega}
[{\bf p},{\bf x}]\right\}\nonumber\\
&&\times\left[1-\frac{1}{\hbar}\int_0^{\hbar\beta}d\tau\,V_{\rm int}({\bf p}(\tau),{\bf x}(\tau))
+\frac{1}{2!\hbar^2}\int_0^{\hbar\beta}d\tau_1\int_0^{\hbar\beta}d\tau_2\,
V_{\rm int}({\bf p}(\tau_1),{\bf x}(\tau_1))V_{\rm int}({\bf p}(\tau_2),{\bf x}(\tau_2))-
\ldots\right].
\end{eqnarray}
Defining harmonic expectation values with respect to the restricted path integral as
\begin{equation}
\label{vpt07}
\meanrp{\ldots}=
\frac{(2\pi\hbar)^3}{Z^{{\bf p}_0,{\bf x}_0}_{\bf \Omega}}\oint{\cal D'}^3x {\cal D}^3p
\,\ldots\,\delta({\bf x}_0-\overline{{\bf x}(\tau)})\delta({\bf p}_0-\overline{{\bf p}(\tau)})
\exp\left\{-\frac{1}{\hbar}{\cal A}^{{\bf p}_0,{\bf x}_0}_{\bf \Omega}[{\bf p},{\bf x}]
\right\},
\end{equation}
the perturbation expansion for the partition function (\ref{vpt06}) reads
\begin{equation}
\label{vpt08}
Z^{{\bf p}_0,{\bf x}_0}=Z^{{\bf p}_0,{\bf x}_0}_{\bf \Omega}\,\sum\limits_{n=0}^\infty
\frac{(-1)^n}{\hbar^nn!}\,
\meanrp{\left(\int_0^{\hbar\beta}d\tau\,V_{\rm int}({\bf p}(\tau),{\bf x}(\tau))\right)^n}.
\end{equation}
This power series expansion can be rewritten in the exponential form
\begin{equation}
\label{vpt09}
Z^{{\bf p}_0,{\bf x}_0}=Z^{{\bf p}_0,{\bf x}_0}_{\bf \Omega}\,\exp\left\{
\sum\limits_{n=0}^\infty
\frac{(-1)^n}{\hbar^nn!}\,
\cumrp{\left(\int_0^{\hbar\beta}d\tau\,V_{\rm int}({\bf p}(\tau),{\bf x}(\tau))\right)^n}\right\},
\end{equation}
where the subscript $c$ on the expectation values indicates cumulants. 
The lowest cumulants are related 
to the full expectation values as follows:
\begin{eqnarray}
\label{vpt10}
\cumrp{O_1({\bf p}(\tau_1),{\bf x}(\tau_1))}&=&\meanrp{O_1({\bf p}(\tau_1),{\bf x}(\tau_1))},
\nonumber\\
\cumrp{O_1({\bf p}(\tau_1),{\bf x}(\tau_1))O_2({\bf p}(\tau_2),{\bf x}(\tau_2))}
&=&\meanrp{O_1({\bf p}(\tau_1),{\bf x}(\tau_1))O_2({\bf p}(\tau_2),{\bf x}(\tau_2))}\nonumber\\
&&-\meanrp{O_1({\bf p}(\tau_1),{\bf x}(\tau_1))}
\meanrp{O_2({\bf p}(\tau_2),{\bf x}(\tau_2))},\nonumber\\
&\vdots&\quad,
\end{eqnarray}
where $O_i({\bf p}(\tau_j),{\bf x}(\tau_j))$ denotes any observable depending on 
position and momentum.
Recalling the relation (\ref{ham12}) between partition function (\ref{vpt09}) 
and effective classical 
Hamiltonian $H_{\rm eff}({\bf p}_0,{\bf x}_0)$, we obtain
from (\ref{vpt09}) the 
effective classical Hamiltonian as a cumulant expansion:
\begin{equation}
\label{vpt11}
H_{\rm eff}({\bf p}_0,{\bf x}_0)=-\frac{1}{\beta}{\rm ln}\,Z^{{\bf p}_0,{\bf x}_0}_{\bf \Omega}
+\frac{1}{\beta}\sum\limits_{n=1}^\infty
\frac{(-1)^n}{\hbar^nn!}\,
\cumrp{\left(\int_0^{\hbar\beta}d\tau\,V_{\rm int}({\bf p}(\tau),{\bf x}(\tau))\right)^n}.
\end{equation}
Up to now,
we did not make any approximation. The expansion on the right-hand side 
is an exact expression for the effective classical Hamiltonian
for any ${\bf \Omega}$. 

For systems with a nontrivial interaction, we are capable of calculating
only some initial truncated part of the series (\ref{vpt11}), say up to the $N$th 
order, leading to the approximate effective classical Hamiltonian
\begin{equation}
\label{vpt12}
{\cal H}_{\bf \Omega}^{(N)}({\bf p}_0,{\bf x}_0)=-
\frac{1}{\beta}{\rm ln}\,Z^{{\bf p}_0,{\bf x}_0}_{\bf \Omega}
+\frac{1}{\beta}\sum\limits_{n=1}^N
\frac{(-1)^n}{\hbar^nn!}\,
\cumrp{\left(\int_0^{\hbar\beta}d\tau\,V_{\rm int}({\bf p}(\tau),{\bf x}(\tau))\right)^n}.
\end{equation}
This 
depends explicitly on the three parameters ${\bf \Omega}$.
Since the exact expression (\ref{vpt11}) is independent of ${\bf \Omega}$,
the best approximation for ${\cal H}_{\bf \Omega}^{(N)}({\bf p}_0,{\bf x}_0)$ should depend
on ${\bf \Omega}$ {\it minimally}. Thus the optimal solution will be found
by determining the parameters from the conditions
\begin{equation}
\label{vpt13}
\nabla_{\bf \Omega} {\cal H}_{\bf \Omega}^{(N)}({\bf p}_0,{\bf x}_0)\stackrel{!}{=}0.
\end{equation}
Let us denote the optimal variational parameters to $N$th order by
\begin{equation}
\label{vpt14}
{\bf \Omega}^{(N)}=\left(\osa^{(N)}({\bf p}_0,{\bf x}_0), 
\osb^{(N)}({\bf p}_0,{\bf x}_0), \op^{(N)}({\bf p}_0,{\bf x}_0)\right). 
\end{equation}
Inserting these
into Eq.~(\ref{vpt12}) yields the optimal effective classical Hamiltonian
${\cal H}^{(N)}({\bf p}_0,{\bf x}_0)$.
\subsection{First-Order Effective Classical Potential}
The first-order approximation 
of the effective classical Hamiltonian (\ref{vpt12}) reads
\begin{equation}
\label{fir00}
{\cal H}_{\bf \Omega}^{(1)}({\bf p}_0,{\bf x}_0)=-
\frac{1}{\beta}{\rm ln}\,Z^{{\bf p}_0,{\bf x}_0}_{\bf \Omega}
-\meanrp{V_{\rm int}({\bf p},{\bf x})}.
\end{equation}
In writing the last term we have used the fact that, as a consequence of the time 
translation invariance of the system, the first-order expectation 
value of $V_{\rm int}({\bf x})$ is independent
of the euclidean time $\tau$.

In order to calculate ${\cal H}_{\bf \Omega}^{(1)}({\bf p}_0,{\bf x}_0)$, we use
the two-point correlation functions derived in App.~\ref{appgenrp}, and
the vanishing of the linear expectations, e.g.
\begin{equation}
\label{fir01}
\meanrp{p_x(\tau)-{p_0}_x}=0
\end{equation}
to find
\begin{equation}
\label{fir02}
{\cal H}_{\bf \Omega}^{(1)}({\bf p}_0,{\bf x}_0)=
\frac{{\bf p}_0^2}{2M}-\frac{1}{2}\omega_c
l_z({\bf p}_0,{\bf x}_0)+\frac{1}{8}M\omega_c^2(x_0^2+y_0^2)+
W_{\bf \Omega}^{(1)}({\bf x}_0),
\end{equation}
where we have collected all terms depending on the variational parameters ${\bf \Omega}$
in the potential
\begin{equation}
\label{fir03}
W_{\bf \Omega}^{(1)}({\bf x}_0)=-\frac{1}{\beta}{\rm ln}\,
Z^{{\bf p}_0,{\bf x}_0}_{\bf \Omega}
+(\omega_c-\osa)\,b^2_\perp({\bf x}_0)-\frac{1}{4}\left(\osb^2-\omega_c^2 \right)
\,a^2_\perp({\bf x}_0)-\frac{1}{2}M\op^2a^2_\parallel({\bf x}_0)+
\meanrp{V({\bf x})}.
\end{equation}
The quantities $a^2_\perp({\bf x}_0)$ and $a^2_\parallel({\bf x}_0)$ are the transverse
and longitudinal fluctuation widths
\begin{equation}
\label{fir03b}
a^2_\perp({\bf x}_0)=G^{{\bf p}_0,{\bf x}_0}_{xx}(0),
\quad a^2_\parallel({\bf x}_0)=G^{{\bf p}_0,{\bf x}_0}_{zz}(0),
\quad b^2_\perp({\bf x}_0)=G^{{\bf p}_0,{\bf x}_0}_{xp_y}(0).
\end{equation}
Note that the potential (\ref{fir03}) is independent of ${\bf p}_0$. 
This means that the approximation 
(\ref{fir02}) to the effective classical Hamiltonian contains no coupling
of the momentum ${\bf p}_0$ to a variational parameter ${\bf \Omega}$, such that
the optimal ${\bf \Omega}^{(1)}$ determined by 
minimizing ${\cal H}_{\bf \Omega}^{(1)}({\bf p}_0,{\bf x}_0)$ is independent of ${\bf p}_0$.
We may therefore integrate out ${\bf p}_0$ in the phase space representation of the first-order
approximation for the partition function
\begin{equation}
\label{fir04}
Z^{(1)}=\int\frac{d^3x_0d^3p_0}{(2\pi\hbar^3)}\,
e^{-\beta{\cal H}_{\bf \Omega}^{(1)}({\bf p}_0,{\bf x}_0)}
\end{equation}
to find the pure configuration space integral
\begin{equation}
\label{fir05}
Z^{(1)}=\int\frac{d^3x_0}{\lambda_{\rm th}^3}\,
e^{-\beta W_{\bf \Omega}^{(1)}({\bf x}_0)},
\end{equation}
in which  
$W_{\bf \Omega}^{(1)}({\bf x}_0)$ is the  
first-order approximation to the effective classical potential of an electron in 
a potential $V({\bf x})$ and a uniform magnetic field. 
\subsection{Application to the Hydrogen Atom in a Magnetic Field}
We now apply the formulas of the preceding section to the
Hamiltonian (\ref{vpt00}) with an attracting Coulomb potential
\begin{equation}
\label{hyd00}
V({\bf x})=-\frac{e^2}{4\pi\varepsilon_0\,|{\bf x}|}\,,
\end{equation}
where $|{\bf x}|$ is the distance between the electron and the proton.
The only nontrivial problem is the calculation of the expectation value
$\meanrp{V({\bf x}(\tau))}$ in Eq.~(\ref{fir03}).
This is done using the so-called 
{\em smearing formula}, which is a Gaussian convolution of $V({\bf x})$.
This formula was first derived by Feynman and Kleinert~\cite{fk}, and exists now also in
an extension to arbitrary order~\cite{density,hatom}. 
The generalization to position and momentum
dependent observables was given in the phase space formulation~\cite{correlation}. 
We briefly rederive
the first-order smearing formula. The expectation value is defined by
\begin{equation}
\label{hyd01}
\meanrp{V({\bf x}(\tau'))}=\frac{(2\pi\hbar)^3}{Z^{{\bf p}_0,{\bf x}_0}_{\bf \Omega}}
\oint {\cal D'}^3x{\cal D}^3p\,V({\bf x}(\tau'))\,\delta({\bf x}_0-\overline{{\bf x}(\tau)})
\delta({\bf p}_0-\overline{{\bf p}(\tau)})
e^{-{\cal A}^{{\bf p}_0,{\bf x}_0}_{\bf \Omega}[{\bf p},{\bf x}]/\hbar},
\end{equation}
Now we substitute the potential by the expression
\begin{equation}
\label{hyd02}
V({\bf x}(\tau'))=\int d^3x\,V({\bf x})\delta({\bf x}-{\bf x}(\tau'))
=\int d^3x\,V({\bf x})\int\,\frac{d^3\kappa}{(2\pi)^3}\,\exp\left[i\bold{\kappa}\cdot
({\bf x}-{\bf x}_0) \right]
\exp\left\{-\frac{1}{\hbar}
\int_0^{\hbar\beta}d\tau\,{\bf j}(\tau)\cdot[{\bf x}(\tau)-{\bf x}_0]
\right\},
\end{equation}
where we have introduced the source
\begin{equation}
\label{hyd03}
{\bf j}(\tau)=i\hbar\bold{\kappa}\delta(\tau-\tau').
\end{equation}
Inserting the expression (\ref{hyd02}) into Eq.~(\ref{hyd01})
we obtain
\begin{equation}
\label{hyd04}
\meanrp{V({\bf x}(\tau'))}=\frac{1}{Z^{{\bf p}_0,{\bf x}_0}_{\bf \Omega}}
\int d^3x\,V({\bf x})\int\,\frac{d^3\kappa}{(2\pi)^3}\,\exp\left[i\bold{\kappa}\cdot
({\bf x}-{\bf x}_0) \right]\,Z^{{\bf p}_0,{\bf x}_0}_{\bf \Omega}[{\bf j}],
\end{equation}
with the harmonic generating functional
\begin{equation}
\label{hyd05}
Z^{{\bf p}_0,{\bf x}_0}_{\bf \Omega}[{\bf j}]=
(2\pi\hbar)^3\oint {\cal D'}^3x{\cal D}^3p\,\delta({\bf x}_0-\overline{{\bf x}(\tau)})
\delta({\bf p}_0-\overline{{\bf p}(\tau)})
\exp\left\{-\frac{1}{\hbar}{\cal A}^{{\bf p}_0,{\bf x}_0}_{\bf \Omega}[{\bf p},{\bf x}]
-\frac{1}{\hbar}\int_0^{\hbar\beta}d\tau\,{\bf j}(\tau)\cdot[{\bf x}(\tau)-{\bf x}_0]\right\}.
\end{equation}
The solution is 
\begin{equation}
\label{hyd05b}
Z^{{\bf p}_0,{\bf x}_0}_{\bf \Omega}[{\bf j}]=
Z^{{\bf p}_0,{\bf x}_0}_{\bf \Omega}
\exp\left[\frac{1}{2\hbar^2}\int_0^{\hbar\beta}d\tau\int_0^{\hbar\beta}d\tau'\,
{\bf j}(\tau)\,{\bf G}^{{\bf x}_0}(\tau,\tau')\,{\bf j}(\tau')\right]
\end{equation}
with the $3\times 3$-matrix of Green functions of Eq.~(\ref{hyd05c}). The properties
of the Green functions are discussed
in the Appendices~\ref{appgen} and ~\ref{appprop}. Expressing the source
${\bf j}(\tau)$ in terms of $\bold{\kappa}$ via Eq.~(\ref{hyd03}) and performing
the $\tau$-integrations, we arrive at
\begin{equation}
\label{hyd06}
\meanrp{V({\bf r}(\tau'))}=\int d^3x\,V({\bf x})\int\,\frac{d^3\kappa}{(2\pi)^3}\,
\exp\left\{i\bold{\kappa}\cdot[{\bf x}-{\bf x}_0] \right\}\,
\exp\left[-\frac{1}{2}\bold{\kappa}\,{\bf G}^{{\bf x}_0}(0)\,\bold{\kappa} \right].
\end{equation}  
Recognizing that $G^{{\bf x}_0}_{yx}(0)=G^{{\bf x}_0}_{xy}(0)$ vanish, 
the $\bold{\kappa}$-integral is easily calculated and leads to the first-order smearing formula
for an arbitrary position dependent potential
\begin{equation}
\label{hyd07}
\meanrp{V({\bf x}(\tau'))}=\frac{1}{(2\pi)^{3/2}\,a^2_\perp({\bf x}_0)
\sqrt{a^2_\parallel({\bf x}_0)}}\int d^3x
\,V({\bf x})\,
\exp\left[-\frac{(x-x_0)^2+(y-y_0)^2}{2a^2_\perp({\bf x}_0)}
-\frac{(z-z_0)^2}{2a^2_\parallel({\bf x}_0)} \right],
\end{equation}
the right-hand side containing the Gaussian fluctuation widths (\ref{fir03b}).

For the Coulomb potential (\ref{hyd00}) that we are interested in, 
the integral in the smearing formula (\ref{hyd07}) can not be done exactly. An integral
representation for a simple numerical treatment is
\begin{eqnarray}
\label{hyd08}
\meanrp{-\frac{e^2}{4\pi\varepsilon_0\,|{\bf x}|}}&=&-\frac{e^2}{4\pi\varepsilon_0}
\sqrt{\frac{2}{\pi}\,a^2_\parallel({\bf x}_0)}\int\limits_0^1\frac{d\xi}
{a^2_\parallel({\bf x}_0)+\xi^2[a^2_\perp({\bf x}_0)-a^2_\parallel({\bf x}_0)]}\nonumber\\
&&\times\exp\left\{-\frac{\xi^2}{2}\left(\frac{x_0^2+y_0^2}{a^2_\parallel({\bf x}_0)+
\xi^2[a^2_\perp({\bf x}_0)-a^2_\parallel({\bf x}_0)]}+\frac{z_0^2}{a^2_\parallel({\bf x}_0)} 
\right) \right\}.
\end{eqnarray}
With this expression we know the entire first-order effective
classical potential (\ref{fir03}) for an electron in a Coulomb potential and a 
uniform magnetic field which has to be optimized in the variational
parameters ${\bf \Omega}$.
\section{Results}
\label{res}
We are now going to optimize the effective classical 
potential by extremizing it in ${\bf \Omega}$ at different 
temperatures and magnetic field strengths. In the 
zero-temperature
limit this will produce the ground state energy.
\subsection{Effective Classical Potential for Different Temperatures and Magnetic Field
Strengths}
\begin{figure}
\centerline{
\epsfxsize=12cm \epsfbox{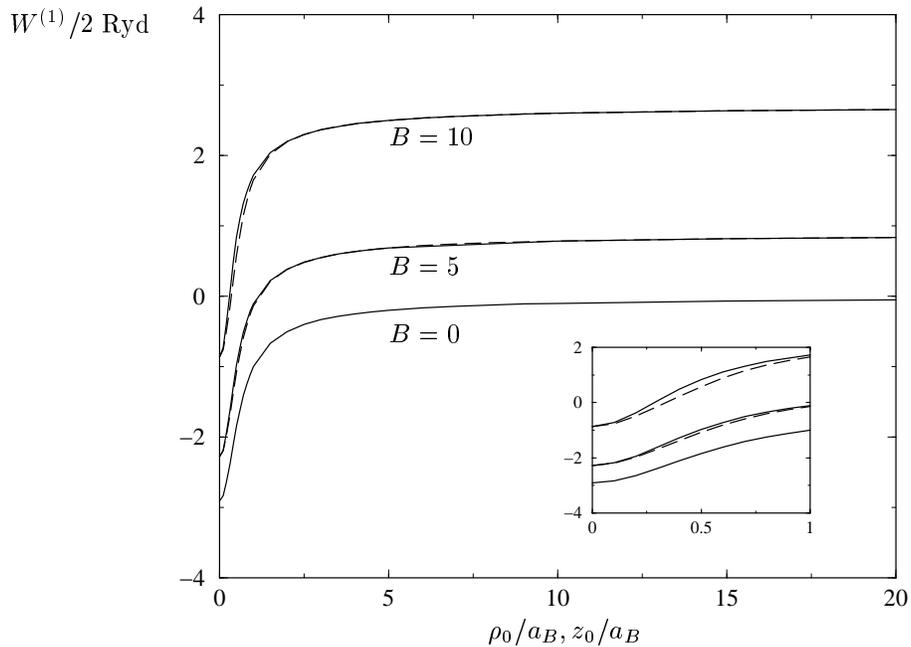}
}
\caption{\label{wofq1} Effective classical potential as a function of the coordinate 
$\rho_0=\sqrt{x_0^2+y_0^2}$ perpendicular to the field lines at $z_0=0$ (solid curves),
and parallel to the magnetic field as a function of $z_0$ at $\rho_0=0$ (dashed curves).
The inverse
temperature is fixed at $\beta=1$, and the strengths of the magnetic field $B$ are varied
(all in natural units). The small 
figure enlarges the range $0\le \rho_0,z_0\le1$ with noticeable anisotropy.}
\end{figure}

\begin{figure}
\centerline{
\epsfxsize=12cm \epsfbox{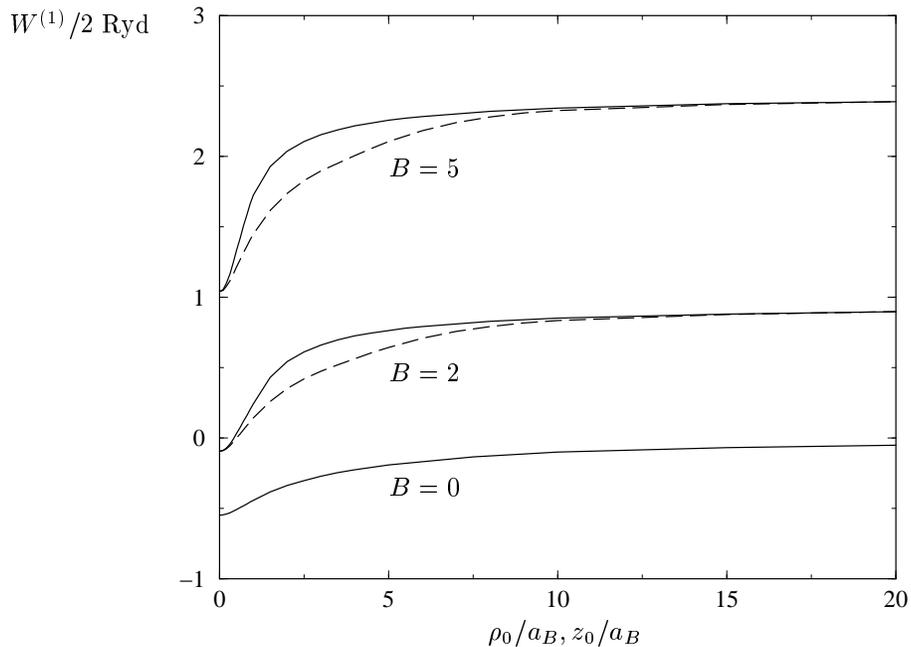}
}
\caption{\label{wofq100} 
Analogous plot to Fig.~\ref{wofq1}, but at the larger inverse temperature $\beta=100$.
}
\end{figure}

The optimization of $W_{{\bf \Omega}}^{(1)}({\bf x}_0)$ proceeds by minimization
in ${\bf \Omega}$ and must be done for each value of 
${\bf x}_0$. Reinserting the optimal parameters ${\bf \Omega}^{(1)}({\bf x}_0)$ into 
the expressions (\ref{fir03}) and (\ref{hyd08}), we obtain the optimal first-order effective
classical potential $W^{(1)}({\bf x}_0)$. The calculations
are done numerically, where we used natural units $\hbar=e^2/4\pi\varepsilon_0=k_B=c=M=1$.
This means that energies are measured in units of 
$\epsilon_0=Me^4/(4\pi\varepsilon_0)^2\hbar^2\equiv 2\,{\rm Ryd}\approx 27.21\,{\rm eV}$, 
temperatures in
$\epsilon_0/k_B\approx 3.16\times 10^5\,{\rm K}$, distances in Bohr radii 
$a_B=(4\pi\varepsilon_0)^2\hbar^2/Me^2\approx
0.53\times 10^{-10}\,{\rm m}$, and magnetic field strengths in 
$B_0=e^3M^2/\hbar^3(4\pi\varepsilon_0)^2\approx 2.35\times 10^5\,{\rm T}=2.35\times10^9\,
{\rm G}$. 
Figure~\ref{wofq1} shows the resulting curves for various magnetic field strengths $B$ and
an inverse tempature $\beta=1/T=1$. Examples of the lower temperature behaviour 
are shown in Fig.~\ref{wofq100} for $\beta=100$. 
To see the expected anisotropy of the curves in the magnetic field direction
and in the plane perpendicular to it, we plot simultanously the curves for 
$W^{(1)}({\bf x}_0)$ transversal to the magnetic field as a function of 
$\rho_0=\sqrt{x_0^2+y_0^2}$ at $z=0$ (solid curves) and parallel as a function
of $z_0$ at $\rho_0=0$ (dashed curves). The curves become strongly anisotropic
for low temperatures and increasing field
strengths (Fig.~\ref{wofq100}).
At a given field strength $B$, the two curves converge for large
distances from the origin, where the proton resides, to the same constant depending on $B$. 
This
is due to the decreasing influence of the Coulomb interaction which shows the classical
$1/r$-behaviour in each direction. When approaching the classical high-temperature
limit, the effect of anisotropy becomes less important since the violent thermal 
fluctuations do not have a preferred direction (see Fig.~\ref{wofq1}). For 
$\rho_0\to\infty$ or $z_0\to\infty$, the expectation value of the Coulomb potential 
(\ref{hyd08}) tends to zero. The remaining effective classical potential
\begin{equation}
\label{res01}
W_{\bf \Omega}^{(1)}({\bf x}_0)\longrightarrow -\frac{1}{\beta}{\rm ln}\,
Z^{{\bf p}_0,{\bf x}_0}_{\bf \Omega}
+(\omega_c-\osa)\,b^2_\perp-\frac{1}{4}\left(\osb^2-\omega_c^2 \right)
\,a^2_\perp-\frac{1}{2}M\op^2a^2_\parallel
\end{equation}
is a constant with regard to the position ${\bf x}_0$,
and the optimization yields $\osa^{(1)}=\osb^{(1)}=\omega_c$
and $\op^{(1)}=0$, leading to the asymptotic constant value
\begin{equation}
\label{res02}
W^{(1)}({\bf x}_0)\longrightarrow -\frac{1}{\beta}{\rm ln}\,
\frac{\hbar\beta\omega_c/2}{\sinh{\hbar\beta\omega_c/2}}.
\end{equation} 
The $B=0$~-curves are of course identical with those obtained from variational
perturbation theory for the hydrogen atom~\cite{hatom}. 
\subsection{Ground State Energy of the Hydrogen Atom in Uniform Magnetic Field}
\label{vptlow}
In what follows we investigate the zero-temperature behaviour of the theory.  
Figures~\ref{wofq1} and \ref{wofq100} show that the minimum of each
potential curve lies at the origin. This means that the first-order approximation to the
ground state energy for a fixed
magnitude of the magnetic field $B$ is found by considering the zero-temperature limit of
the first-order effective classical potential in the origin
\begin{equation}
\label{gr00} 
E^{(1)}=\lim_{\beta\to\infty}W^{(1)}(0).
\end{equation}
Thus we obtain from Eq.~(\ref{fir03}) the variational expression for the ground state energy:
\begin{equation}
\label{gr01}
E^{(1)}_{\bf \Omega}(B)=\frac{\hbar}{4\osb}\left(\osb^2+\omega_c^2\right)+\frac{\hbar\op}{4}-
\frac{e^4}{4\pi\varepsilon_0}\meanzero{\frac{1}{|{\bf x}|}},
\end{equation}
where the expectation value for the Coulomb potential (\ref{hyd08}) can now be calculated
exactly since the exponential in the integral simplifies to unity:
\begin{equation}
\label{gr02}
\meanzero{\frac{1}{|{\bf x}|}}=2\sqrt{\frac{M}{\pi\hbar}}\times\left\{
\begin{array}{cl}  
\DS\sqrt{\frac{\op\osb}{\op-\osb}}\,{\rm arctan}\,\sqrt{\frac{2\op}{\osb}-1}, & \quad 2\op > \osb,\\
\DS\sqrt{\op}, & \quad 2\op = \osb,\\
\DS\frac{1}{2i}\sqrt{\frac{\op\osb}{\op-\osb}}\,{\rm ln}\,
\frac{1+i\sqrt{2\op/\osb-1}}{1-i\sqrt{2\op/\osb-1}}, & \quad 2\op < \osb.
\end{array}\right.
\end{equation}
The equations (\ref{gr01}) and (\ref{gr02}) are independent of the frequency parameter
$\osa$ such that the optimization of the first-order expression for the ground state energy 
(\ref{gr01}) requires the satisfying of the equations
\begin{equation}
\label{gr03}
\frac{\partial E^{(1)}_{\bf \Omega}(B)}{\partial \osb}\stackrel{!}{=}0,\qquad
\frac{\partial E^{(1)}_{\bf \Omega}(B)}{\partial \op}\stackrel{!}{=}0.
\end{equation}
Reinserting the resulting values $\osb^{(1)}$ and $\op^{(1)}$ into Eq.~(\ref{gr01}) 
yields the first-order approximation for the ground state energy $E^{(1)}(B)$. In the absence of 
the Coulomb interaction the optimization with respect to $\osb$ yields $\osb^{(1)}=\omega_c$,
rendering the ground state energy $E^{(1)}(B)=\omega_c/2$, which
is the zeroth Landau level. An optimal value for $\op$ does not exist since  
the dependece of the ground state energy of this parameter is linear in Eq.~(\ref{gr01}) in this
special case. To obtain the lowest energy, this parameter can be set to zero 
(all optimal frequency parameters used in the optimization procedure turn out to be 
nonnegative). 
For a vanishing magnetic field, $B=0$, Eq.~(\ref{gr01}) exactly reproduces the first-order
variational result for the ground state energy of the hydrogen atom, 
$E^{(1)}(B=0)\approx -0.42\, {\rm [2\,Ryd]}$, obtained in Ref.~\cite{hatom}.

\begin{figure}
\centerline{\epsfxsize=12cm \epsfbox{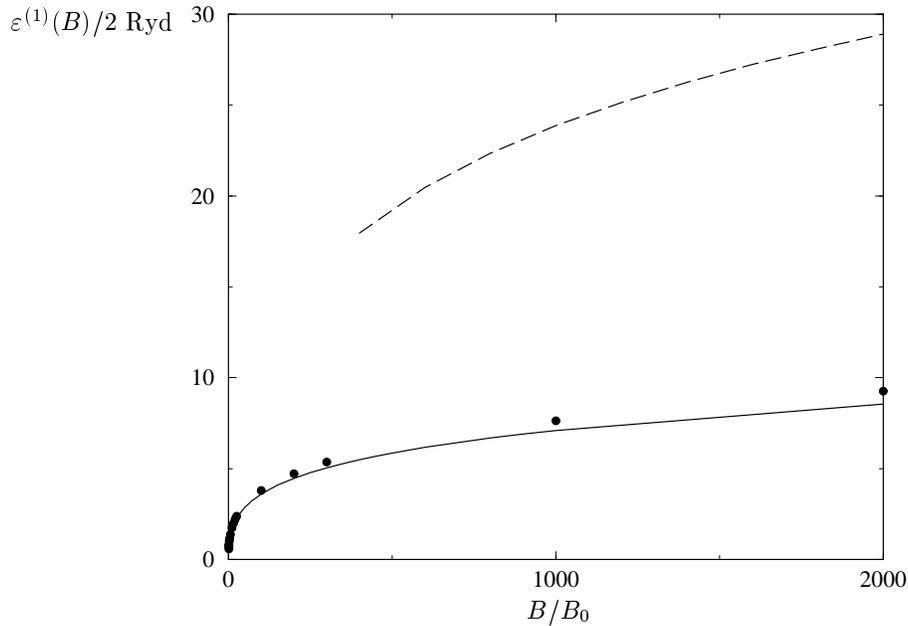}}
\caption[]{\label{bofb} First-order variational result for the binding energy 
as a function of the
strength of the magnetic field. The dots indicate the values of Ref.~\cite{zinn-justin}. 
The dashed curve shows the simple estimate of Landau-Lifschitz~\cite{landau} 
$0.5\, {\rm ln}^2B$, which is closely related to the ground state energy
of the one-dimensional hydrogen atom~\cite{loudon,haines}.}
\end{figure}

To investigate the asymptotics in the strong-field limit $B\to\infty$, it is useful
to extract the leading term $\omega_c/2$. Thus we define the binding energy
\begin{equation}
\label{gr04}
\varepsilon(B)\equiv\frac{\omega_c}{2}-E(B)
\end{equation}
which possesses an characteristic strong-field behaviour to be discussed in detail 
subsequently. The result 
is shown in Fig.~\ref{bofb} as a function of the
magnitude of the magnetic field $B$, where it is compared with the high-accuracy results of
Ref.~\cite{zinn-justin}. As a first-order approximation, this result is satisfactory.
It is of the same quality like other first-order results, for example those from the operator
optimization method in first order of Ref.~\cite{feranshuk}. The advantage
of variational perturbation theory is that it yields good results over the complete range
of the coupling strength, here the magnetic field. Moreover, as a consequence of
the exponential convergence~\cite[Chap. 5]{PI}, higher orders
of variational perturbation theory push the approximative result of any quantity very rapidly 
towards the exact value.
\subsubsection{The Weak-Field Case}
We investigate now the weak-field behaviour of our theory starting from the expression
(\ref{gr04}) and the expectation value of the Coulomb potential (\ref{gr02}) in natural
units:
\begin{equation}
\label{we00}
\varepsilon^{(1)}_{\eta,\Omega}(B)=
\frac{B}{2}-\frac{\Omega}{4}\left(1+\frac{\eta}{2}\right)-\frac{B^2}{4\Omega}
-\sqrt{\frac{\eta\Omega}{2\pi}}h(\eta)
\end{equation}
with
\begin{equation}
\label{we01}
h(\eta)=\frac{1}{\sqrt{1-\eta}}\,{\rm ln}\,\frac{1-\sqrt{1-\eta}}{1+\sqrt{1-\eta}}.
\end{equation}
In comparison with Eq.~(\ref{gr01}) we introduced new variational parameters
\begin{equation}
\label{we02}
\eta\equiv\frac{2\op}{\osb},\qquad \Omega\equiv\osb
\end{equation}
and utilized, as the calculations for the binding energy showed, that always $\eta\le 1$.
Performing the derivatives with respect to these variational parameters and setting them zero
yields conditional equations which can be written after some manipulations as
\begin{eqnarray}
\label{we03}
\frac{\Omega}{8}+\sqrt{\frac{\Omega}{2\pi\eta}}\frac{1}{1-\eta}\left(1+\frac{1}{2}
\frac{1}{\sqrt{1-\eta}}{\rm ln}\,\frac{1-\sqrt{1-\eta}}{1+\sqrt{1-\eta}}\right)&\stackrel{!}{=}&0,
\nonumber\\
\label{we04}
\frac{1}{4}+\frac{\eta}{8}-\frac{B^2}{4\Omega^2}+\frac{1}{2}\sqrt{\frac{\eta}{2\pi\Omega}}
\frac{1}{\sqrt{1-\eta}}{\rm ln}\,\frac{1-\sqrt{1-\eta}}{1+\sqrt{1-\eta}}&\stackrel{!}{=}&0.
\end{eqnarray}
Expanding the variational parameters into perturbation series of the square magnetic field
$B^2$,
\begin{equation}
\label{we05}
\eta(B)=\sum\limits_{n=0}^{\infty}\,\eta_n B^{2n},\qquad 
\Omega(B)=\sum\limits_{n=0}^{\infty}\,\Omega_n B^{2n}
\end{equation}
and inserting these expansions into the self-consistency conditions (\ref{we03}) and 
(\ref{we04}) we obtain order by order the coefficients given in Table~\ref{tab1}.
Inserting these values into the expression for the binding energy (\ref{we00}) and expand
with respect to $B^2$, we obtain the perturbation series
\begin{equation}
\label{we06}
\varepsilon^{(1)}(B)=\frac{B}{2}-\sum\limits_{n=0}^\infty\,\varepsilon_n B^{2n}.
\end{equation} 
The first coefficients are also given in Table~\ref{tab1}. We find thus the important 
result that the first-order variational perturbation solution possesses a perturbative
behaviour with respect to the square magnetic field strength $B^2$ in the weak-field limit thus
yielding the correct asymptotics. The coefficients differ in higher order from the exact ones
but are improved in higher orders of the variational perturbation 
theory~\cite[Chap. 5]{PI}. 
\begin{table}
\caption[]{\label{tab1} Perturbation coefficients up to order $B^6$ for the weak-field
expansions of the variational parameters and the binding energy in comparison to the
exact ones of Ref.~\cite{cizek1}.}
\begin{tabular}{c|cccc}
$n$ & 0 & 1 & 2 & 3\\ \hline\hline
$\eta_n$ & $1.0$ & $\DS-\frac{405\pi^2}{7168}\approx -0.5576$ & 
$\DS\frac{16828965\pi^4}{1258815488}\approx 1.3023$ &
$\DS-\frac{3886999332075\pi^6}{884272562962432}\approx -4.2260$\\[1mm] \hline
$\Omega_n$ & $\DS\frac{32}{9\pi}\approx 1.1318$ & $\DS\frac{99\pi}{224}\approx 1.3885$ &
$\DS -\frac{1293975\pi^3}{19668992}\approx -2.03982$ & 
$\DS\frac{524431667187\pi^5}{27633517592576}\approx 5.8077$
\\[1mm] \hline
$\varepsilon_n$ & $\DS -\frac{4}{3\pi}\approx -0.4244$ & $\DS\frac{9\pi}{128}\approx 0.2209$ & 
$\DS -\frac{8019\pi^3}{1835008}\approx -0.1355$ & $\DS\frac{256449807\pi^5}{322256764928}
\approx 0.2435$\\[1mm] \hline
$\varepsilon_n$~\cite{cizek1} & $-0.5$ & $0.25$ & $\DS -\frac{53}{192}\approx -0.2760$ &
$\DS\frac{5581}{4608}\approx 1.2112$ 
\end{tabular}
\end{table}
\subsubsection{Asymptotical Behaviour in the Strong-Field Regime}
In the discussion of the pure magnetic field below Eq.~(\ref{gr03}) we have mentioned 
that the variational calculation for the ground state energy which is thus associated 
with the zeroth Landau level yields a frequency $\osb\propto B$ while $\op=0$. Therefore
we use the assumption (with $\os\equiv \osb$)
\begin{equation}
\label{st00}
\os \gg 2\op,\qquad \op \ll B
\end{equation}
for the consideration of the ground state energy (\ref{gr01}) of the hydrogen atom in 
a strong magnetic field. In a first step we expand the last expression of the 
expectation value (\ref{gr02}) which corresponds to the condition (\ref{st00}) 
in terms of $2\op/\os$ and reinsert this expansion in the equation of the ground state 
energy (\ref{gr01}). 
Then we omit all terms proportional to $C/\os$ where $C$ stands for any expression
with a value much smaller than the field strength $B$. In natural units, we thus 
obtain the strong-field approximation for the first-order binding energy (\ref{gr04})
\begin{equation}
\label{st01}
\varepsilon_{\os,\op}^{(1)}=\frac{B}{2}-\left(\frac{\os}{4}+\frac{B^2}{4\os}+\frac{\op}{4}
+\sqrt{\frac{\op}{\pi}}\,{\rm ln}\,\frac{\op}{2\os}\right).
\end{equation}
As usual, we consider the zeros of the derivatives with respect to the variational parameters
\begin{equation}
\label{st02}
\frac{\partial \varepsilon_{\os,\op}^{(1)}}{\partial \op}\stackrel{!}{=}0,\qquad
\frac{\partial \varepsilon_{\os,\op}^{(1)}}{\partial \os}\stackrel{!}{=}0,
\end{equation}
which lead to the self-consistence equations
\begin{eqnarray}
\label{st03}
\sqrt{\op}&=&-\frac{2}{\sqrt{\pi}}\left({\rm ln}\,\op-{\rm ln}\,\os+2-{\rm ln}\,2 \right),\\
\label{st04}
\os&=&2\sqrt{\frac{\op}{\pi}}+B\sqrt{1+4\frac{\op}{\pi B^2}}
\end{eqnarray}
Let us first consider the last equation. Utilizing the second of the conditions (\ref{st00})
we expand the second root around unity yielding the expression
\begin{equation}
\label{st05}
\os=B+2\sqrt{\frac{\op}{\pi}}+2 \frac{\op}{\pi B}-4\frac{\op^2}{\pi^2 B^3}+\ldots,
\end{equation} 
where the terms are sorted with regard to their contribution starting with the biggest. Since
we are interested in the strong $B$ limit, we can obviously neglect terms suppressed
by powers of $1/B$. Thus we only consider the following terms for the moment:
\begin{equation}
\label{st06}
\os\approx B+2\sqrt{\frac{\op}{\pi}}.
\end{equation}
Inserting this into the other condition (\ref{st03}), expanding the corresponding logarithm,
and, once more, neglecting terms of order $1/B$, we find
\begin{equation}
\label{st07}
\sqrt{\op}\approx\frac{2}{\sqrt{\pi}}\left({\rm ln}\,B-{\rm ln}\,\op^{(1)}+{\rm ln}\,2-2 \right).
\end{equation} 
To obtain a tractable approximation for $\op$, we perform some iterations starting from 
\begin{equation}
\label{st08}
\sqrt{\op^{(1)}}=\frac{2}{\sqrt{\pi}}{\rm ln}\,2Be^{-2}
\end{equation}
Reinserting this on the right-hand side of Eq.~(\ref{st07}), one obtains the second iteration
$\sqrt{\op^{(2)}}$. We stop this procedure after an additional reinsertion which yields
\begin{equation}
\label{st09}
\sqrt{\op^{(3)}}=\frac{2}{\sqrt{\pi}}\left({\rm ln}\,2Be^{-2}-2{\rm ln}\left[\frac{2}{\sqrt{\pi}}
\left\{{\rm ln}\,2Be^{-2}-2{\rm ln}\,\left(\frac{2}{\sqrt{\pi}}{\rm ln}\,2Be^{-2}\right) 
\right\}\right] \right).
\end{equation} 
The reader may convince himself that this iteration procedure indeed converges. 
For a subsequent systematical extraction of terms essentially contributing to the binding
energy, the expression (\ref{st09}) is not satisfactory. Therefore it is better to separate
the leading term in the curly brackets and expand the logarithm of the remainder. Then this 
proceeding is applied to the expression in the angular brackets and so on. Neglecting terms 
of order ${\rm ln}^{-3} B$, we obtain
\begin{equation}
\label{st10}
\sqrt{\op^{(3)}}\approx \frac{2}{\sqrt{\pi}}\left({\rm ln}\,2Be^{-2}+{\rm ln}\frac{\pi}{4} 
-2{\rm ln}{\rm ln}\,2Be^{-2}\right).
\end{equation}
The double-logarithmic term can be expanded in a similar way as described above:
\begin{equation}
\label{st11}
{\rm ln}{\rm ln}\,2Be^{-2}={\rm ln}\left[{\rm ln}\,B\left(1+\frac{{\rm ln}\,2-2}{{\rm ln}\,B} 
\right) \right]= {\rm ln}{\rm ln}\,B+\frac{{\rm ln}\,2-2}{{\rm ln}\,B}-\frac{1}{2}
\frac{({\rm ln}\,2-2)^2}{{\rm ln}^2B}+{\cal O}({\rm ln}^{-3} B).
\end{equation}
Thus the expression (\ref{st10}) may be rewritten as
\begin{equation}
\label{st12}
\sqrt{\op^{(3)}}= \frac{2}{\sqrt{\pi}}\left( {\rm ln}\,B -2{\rm ln}{\rm ln}\,B+
\frac{2a}{{\rm ln}\,B}+\frac{a^2}{{\rm ln}^2B}+b \right) +{\cal O}({\rm ln}^{-3} B)
\end{equation}
with abbreviations 
\begin{equation}
\label{st13}
a=2-{\rm ln}\,2 \approx 1.307,\qquad b= {\rm ln}\frac{\pi}{2}-2\approx -1.548.
\end{equation}
The first observation is that the variational parameter $\op$ is always 
much smaller
than $\os$ in the high $B$-field limit. Thus we can further simplify the approximation
(\ref{st06}) by replacing
\begin{equation}
\label{st14}
\os\approx B\left(1+\frac{2}{B}\sqrt{\frac{\op}{\pi}}\right)\longrightarrow B
\end{equation}
without affecting the following expression for the binding energy. Inserting the solutions
(\ref{st12}) and (\ref{st14}) into the equation for the binding energy (\ref{st01}) and 
expanding the logarithmic term once more as described, we find up to the 
order ${\rm ln}^{-2}B$:
\begin{eqnarray}
\label{st15}
\varepsilon^{(1)}(B)&=&\frac{1}{\pi}\left({\rm ln}^2B -4\,{\rm ln}\,B\; {\rm ln}{\rm ln}\,B
+4\,{\rm ln}^2{\rm ln}\,B -4b\,{\rm ln}{\rm ln}\,B
+2(b+2)\,{\rm ln}\,B
+b^2
-\frac{1}{{\rm ln}\,B}\left[8\,{\rm ln}^2{\rm ln}\,B-8b\,{\rm ln}{\rm ln}\,B+2b^2 
\right]\right)\nonumber\\
&&+{\cal O}({\rm ln}^{-2}B)
\end{eqnarray}
Note that the prefactor $1/\pi$ of the leading ${\rm ln}^2B$-term 
differs from a value $1/2$ obtained by Landau and Lifschitz~\cite{landau}.
Our different value is a consequence of using a harmonic trial system. 
The calculation of higher orders
in variational perturbation theory would improve the value of the prefactor. 

At a magnetic field strength $B=10^5 B_0$, which corresponds to $2.35\times 10^{10}\,{\rm T}=
2.35\times10^{14}\,{\rm G}$, 
the contribution from the first six terms is $22.87\,[2\,{\rm Ryd}]$. 
The next three terms suppressed
by a factor ${\rm ln}^{-1} B$ contribute $-2.29\,[2\,{\rm Ryd}]$, 
while an estimate for the ${\rm ln}^{-2} B$-terms
yields nearly $-0.3\,[2\,{\rm Ryd}]$. Thus we find 
\begin{equation}
\label{st16}
\varepsilon^{(1)}(10^5)=20.58\pm 0.3\,[2\,{\rm Ryd}].
\end{equation} 
This is in very good agreement with the value $20.60\,[2\,{\rm Ryd}]$ 
obtained from the full treatment described in Sec.~\ref{vptlow}.

Table~\ref{asymp} lists the values of the first six terms of Eq.~(\ref{st15}).
This shows in particular the significance of the second-leading term  
$-(4/\pi){\rm ln}\,B\;{\rm ln}{\rm ln}\,B$, which is of the same order of the leading term
$(1/\pi){\rm ln}^2B$ but with an opposite sign. In Fig.~\ref{bofb},  
we have plotted the expression 
\begin{equation}
\label{st17}
\varepsilon_L(B)=\frac{1}{2}\,{\rm ln}^2B
\end{equation} 
from Landau and Lifschitz~\cite{landau} to illustrate that it gives far too large
binding energies even at very large magnetic fields, e.g. at $2000 B_0\propto 10^{12}\,{\rm G}$.

This strength of magnetic field appears on surfaces of neutron stars 
($10^{10}-10^{12}\,{\rm G}$). A recently discovered new type of neuton star is the
so-called magnetar. In these, charged particles such as protons and electrons produced by 
decaying neutrons give rise to the giant magnetic field of $10^{15}\,{\rm G}$.
Magnetic fields of white dwarfs reach only up to $10^6-10^8\,{\rm G}$.   
All these magnetic field strengths are far from realization in
experiments. The strongest magnetic fields ever produced in a laboratory were only of the order 
$10^5\,{\rm G}$, an order of magnitude larger than the fields in sun spots which reach about
$0.4\times10^4\,{\rm G}$. Recall, for comparison, that the earth's
magnetic field has the small value of $0.6\,{\rm G}$.

As we see in Fig.~\ref{bofb}, the nonleading terms in Eq.~(\ref{st15}) 
give important contributions to the asymptotic behaviour even at such large magnetic fields. 
It is an unusual property of the asymptotic behaviour that the absolute value of 
the difference between the Landau-expression (\ref{st17}) and our approximation (\ref{st15})
diverges with increasing magnetic field strengths $B$, only the relative difference
decreases.
\begin{table}
\caption{\label{asymp} Example for the competing leading six terms in Eq.~(\ref{st15})
at $B=10^5B_0\approx 2.35\times 10^{14}\,{\rm G}$.}
\begin{tabular}{cccccc}
$(1/\pi){\rm ln}^2B$ & $-(4/\pi){\rm ln}\,B\;{\rm ln}{\rm ln}\,B$ &
$(4/\pi)\,{\rm ln}^2{\rm ln}\,B$ & 
$-(4b/\pi)\,{\rm ln}{\rm ln}\,B$ & $[2(b+2)/\pi]\,{\rm ln}\,B$ & $b^2/\pi$\\ \hline
$42.1912$ & $-35.8181$ & $7.6019$ & $4.8173$ & $3.3098$ & $0.7632$
\end{tabular}
\end{table}
\section{Summary}
We have calculated the effective classical potential for the hydrogen atom
in a magnetic field. For this we have generalized variational perturbation theory 
to make it applicable to physical systems with uniform external magnetic field.  

The effective classical potential containing the complete
quantum statistical information of the system was determined in first-order
variational perturbation theory. For zero-temperature, it gave the
energy of the system.
Our result consists of a single analytic expression which is quite accurate
at all temperatures and magnetic field strengths.  
\section*{Acknowledgments}
We thank Prof.~J.~\v{C}\'\i\v{z}ek and Dr.~J.~Weniger for useful hints and references 
of a perturbative treatment of the ground state properties of the hydrogen atom in 
a magnetic field. The authors also thank Dr.~J.~Ortner and M.~Steinberg 
for discussions of the finite temperature behaviour of this system. For interesting 
discussions we would also like to thank Prof.~J.T.~Devreese and Prof.~G.~Wunner.

One of us (M.B.) is grateful for support by the Studienstiftung des deutschen Volkes. 
\begin{appendix}
\section{Generating Functional for Particle in Magnetic Field and Harmonic Oscillator
Potential}
\label{appgen}
For the determination of the correlation functions of a system,
we need to know the solution of the two-dimensional generating functional
in the presence of an external source ${\bf j}=(j_x,j_y)$:
\begin{equation}
\label{app00}
Z^{{\bf x}_0}[{\bf j}]=\lambda_{\rm th}^2\oint {\cal D}^2x\,
\delta({\bf x}_0-\overline{{\bf x}(\tau)})\,
e^{-{\cal A}^{{\bf x}_0}[{\bf x};{\bf j}]/\hbar}.
\end{equation}
The action of a particle in a magnetic field in $z$-direction and a harmonic oscillator
reads 
\begin{equation}
\label{app01}
{\cal A}^{{\bf x}_0}[{\bf x};{\bf j}]=\int_0^{\hbar\beta}d\tau\,\left[
\frac{M}{2}\dot{\bf x}^2(\tau)-\frac{i}{2}M\omega([{\bf x}(\tau)-{\bf x}_0]
\times\dot{\bf x}(\tau))_z 
+\frac{1}{2}M\Omega^2[{\bf x}(\tau)-{\bf x}_0]^2+{\bf j}(\tau)\cdot({\bf x}(\tau)-{\bf x}_0)  
\right].
\end{equation}
The position dependent terms are centered around ${\bf x}_0=(x_0,y_0)$, which is the temporal 
average of the path ${\bf x}(\tau)$, and thus equal to the 
zero frequency component of the Fourier path, is
\begin{equation}
\label{app02}
{\bf x}(\tau)={\bf x}_0+\sum\limits_{m=1}^\infty\,\left({\bf x}_me^{i\omega_m\tau}+
{\bf x}^\star_me^{-i\omega_m\tau} \right)
\end{equation}
with the Matsubara frequencies $\omega_m=2\pi m/\hbar\beta$ and complex Fourier coefficients
${\bf x}_m={\bf x}_m^{\rm re}+i{\bf x}_m^{\rm im}$. 
Introducing a similar Fourier decomposition for the current ${\bf j}(\tau)$ with Fourier
components ${\bf j}_m$ and using the orthonormality relation
\begin{equation}
\label{app03}
\frac{1}{\hbar\beta}\int_0^{\hbar\beta}d\tau\,e^{i(\omega_m-\omega_n)\tau}=\delta_{m\,n},
\end{equation}
the generating functional can be written as
\begin{equation}
\label{app04}
Z^{{\bf x}_0}[{\bf j}]=\prod\limits_{m=1}^\infty\left[\int\frac{dx_m^{\rm re}dx_m^{\rm im}
dy_m^{\rm re}dy_m^{\rm im}}{(\pi/M\beta\omega_m^2)^2} 
e^{-{\cal A}_m({\bf x}_m,{\bf x}^\star_m;{\bf j}_m,{\bf j}^\star_m)/\hbar} \right]
\end{equation}
with
\begin{eqnarray}
\label{app05}
{\cal A}_m({\bf x}_m,{\bf x}^\star_m;{\bf j}_m,{\bf j}^\star_m)&=&
\hbar\beta M(\omega_m^2+\Omega^2)
([x_m^{\rm re}]^2+[x_m^{\rm im}]^2+[y_m^{\rm re}]^2+[y_m^{\rm im}]^2)
+2i\hbar\beta M \omega\omega_m(x_m^{\rm re}y_m^{\rm im}-x_m^{\rm im}y_m^{\rm re})\nonumber\\
&&+2\hbar\beta(x_m^{\rm re}{j_x}_m^{\rm re}+x_m^{\rm im}{j_x}_m^{\rm im}+
y_m^{\rm re}{j_y}_m^{\rm re}+y_m^{\rm im}{j_y}_m^{\rm im}).
\end{eqnarray}
Expression (\ref{app04}) is equivalent to the path integral (\ref{app00}) and we obtain
after performing the integrations and retransforming the currents 
\begin{equation}
\label{app06}
{\bf j}_m=\frac{1}{\hbar\beta}\int_0^{\hbar\beta}d\tau\,{\bf j}(\tau)e^{-i\omega_m\tau}
\end{equation}
the resulting generating functional
\begin{equation}
\label{app07}
Z^{{\bf x}_0}[{\bf j}]=Z^{{\bf x}_0}\exp\left\{\frac{1}{2\hbar^2}
\int_0^{\hbar\beta}d\tau \int_0^{\hbar\beta}d\tau'\,
{\bf j}(\tau){\bf G}^{{\bf x}_0}(\tau,\tau'){\bf j}(\tau') \right\}
\end{equation}
with the partition function
\begin{equation}
\label{app08}
Z^{{\bf x}_0}\equiv Z^{{\bf x}_0}[0]=
\prod\limits_{m=1}^\infty\,\frac{\omega_m^4}{\omega^2\omega_m^2+
(\omega_m^2+\Omega^2)^2}
\end{equation}
and the $2\times 2$-matrix of Green functions
\begin{equation}
\label{app09}
{\bf G}^{{\bf x}_0}(\tau,\tau')=\left(\begin{array}{cc}
G_{xx}^{{\bf x}_0}(\tau,\tau') & G_{xy}^{{\bf x}_0}(\tau,\tau')\\
G_{yx}^{{\bf x}_0}(\tau,\tau') & G_{yy}^{{\bf x}_0}(\tau,\tau') \end{array}\right).
\end{equation}
The elements of this matrix are position-position correlation functions what can be easily
proved by applying two functional derivatives with respect to the desired component of the 
current to the functional (\ref{app00}), for example
\begin{equation}
\label{app10}
G_{xx}^{{\bf x}_0}(\tau,\tau')=\meanr{(x(\tau)-x_0)\,(x(\tau')-x_0)}
=\left[\hbar^2\frac{1}{Z^{{\bf x}_0}[{\bf j}]}
\frac{\delta^2}{\delta j_x(\tau)\delta j_x(\tau')}Z^{{\bf x}_0}[{\bf j}]\right]_{{\bf j}=0},
\end{equation}
where we have defined expectation values by
\begin{equation}
\label{app11}
\meanr{\ldots}=\frac{\lambda_{\rm th}^2}
{Z^{{\bf x}_0}}\oint {\cal D}^2x\,\ldots\,\delta({\bf x}_0-\overline{{\bf x}(\tau)})
e^{-{\cal A}^{{\bf x}_0}[{\bf x};0]/\hbar}.
\end{equation}
From the above calculation we find the following expressions for the Green functions
in Fourier space ($0\le\tau,\tau'\le\hbar\beta$):
\begin{eqnarray}
\label{app12a}
G_{xx}^{{\bf x}_0}(\tau,\tau')&=&
\meanr{\tx(\tau)\,\tx(\tau')} = G_{yy}^{{\bf x}_0}(\tau,\tau')=\meanr{\ty(\tau)\,\ty(\tau')}
\nonumber\\
&=&\frac{2}{M\beta}\sum\limits_{m=1}^\infty\,
\frac{\omega_m^2+\Omega^2}{\omega^2\omega_m^2+(\omega_m^2+\Omega^2)^2}\,
e^{-i\omega_m(\tau-\tau')},\\
\label{app12b}
G_{xy}^{{\bf x}_0}(\tau,\tau')&=&\meanr{\tx(\tau)\,\ty(\tau')}=
-G_{yx}^{{\bf x}_0}(\tau,\tau')=-\meanr{\ty(\tau)\,\tx(\tau')}\nonumber\\
&=&\frac{2\omega}{M\beta}\sum\limits_{m=1}^\infty\,
\frac{\omega_m}{\omega^2\omega_m^2+(\omega_m^2+\Omega^2)^2}\,
e^{-i\omega_m(\tau-\tau')},
\end{eqnarray}
where, for simplicity, $\tr(\tau)={\bf x}(\tau)-{\bf x}_0$. It is desirable to find analytical
expressions for the Green functions and the partition function (\ref{app08}). All these 
quantities possess the same dominator which can be decomposed as
\begin{equation}
\label{app13}
\omega^2\omega_m^2+(\omega_m^2+\Omega^2)^2=(\omega_m^2+\Omega_+^2)(\omega_m^2+\Omega_-^2)
\end{equation}
with frequencies
\begin{equation}
\label{app14}
\Omega_\pm(\omega,\Omega)=
\sqrt{\Omega^2+\frac{1}{2}\omega^2\pm\omega\sqrt{\Omega^2+\frac{1}{4}\omega^2}}.
\end{equation}
Therefore the partition function (\ref{app08}) can be split into two products, each of which
known from the harmonic oscillator~\cite[Chap. 5]{PI}:
\begin{equation}
\label{app15}
Z^{{\bf x}_0}=\prod\limits_{m=1}^\infty\left[\frac{\omega_m^2}{\omega_m^2+\Omega_+^2}\right]
\prod\limits_{m=1}^\infty\left[\frac{\omega_m^2}{\omega_m^2+\Omega_-^2}\right]
=\frac{\hbar\beta\Omega_+/2}{\sinh{\hbar\beta\Omega_+/2}}\,
\frac{\hbar\beta\Omega_-/2}{\sinh{\hbar\beta\Omega_-/2}}.
\end{equation}
Now we apply the property (\ref{app13}) to decompose the Green functions (\ref{app12a})
into partial fractions, yielding
\begin{equation}
\label{app16}
G_{xx}^{{\bf x}_0}(\tau,\tau')=G_{yy}^{{\bf x}_0}(\tau,\tau')=\frac{1}{M\beta}
\left(\alpha_1\,\sum\limits_{m=-\infty}^\infty\,\frac{1}{\omega_m^2+\Omega_+^2}\,
e^{-i\omega_m(\tau-\tau')}+\alpha_2\,\sum\limits_{m=-\infty}^\infty\,
\frac{1}{\omega_m^2+\Omega_-^2}\,
e^{-i\omega_m(\tau-\tau')}-\frac{1}{\Omega^2} \right)
\end{equation} 
with coefficients
\begin{equation}
\label{app17}
\alpha_1=\frac{\Omega_+^2-\Omega^2}{\Omega_+^2-\Omega_-^2},\qquad 
\alpha_2=-\frac{\Omega_-^2-\Omega^2}{\Omega_+^2-\Omega_-^2}.
\end{equation}
Following Ref.~\cite[Chap. 3]{PI}, sums of the kind occuring in expression (\ref{app16})
are spectral decompositions of the correlation function for the harmonic oscillator
and can be summed up:
\begin{equation}
\label{app18}
\sum\limits_{m=-\infty}^\infty\,\frac{1}{\omega_m^2+\Omega_\pm^2}\,e^{-i\omega_m(\tau-\tau')}=
\frac{\hbar\beta}{2\Omega_\pm}
\frac{\cosh{\Omega_\pm(|\tau-\tau'|-\hbar\beta/2)}}{\sinh{\hbar\beta\Omega_\pm/2}}.
\end{equation}
Thus, the $xx$- and $yy$-correlation functions can be expressed by
\begin{eqnarray}
\label{app19}
&&G_{xx}^{{\bf x}_0}(\tau,\tau')=G_{yy}^{{\bf x}_0}(\tau,\tau')\nonumber\\
&&\hspace*{5mm}=\frac{1}{M\beta}\left(\frac{\hbar\beta}{2\Omega_+}\,
\frac{\Omega_+^2-\Omega^2}{\Omega_+^2-\Omega_-^2}\,
\frac{\cosh{\Omega_+(|\tau-\tau'|-\hbar\beta/2)}}{\sinh{\hbar\beta\Omega_+/2}} 
-\frac{\hbar\beta}{2\Omega_-}\,
\frac{\Omega_-^2-\Omega^2}{\Omega_+^2-\Omega_-^2}\,
\frac{\cosh{\Omega_-(|\tau-\tau'|-\hbar\beta/2)}}{\sinh{\hbar\beta\Omega_-/2}}
-\frac{1}{\Omega^2}
\right),
\end{eqnarray}
where, from Eq.~(\ref{app14}), $\Omega_\pm=\Omega_\pm(\omega,\Omega)$ are functions
of the original frequencies $\omega$ from the magnetic field and $\Omega$ from the 
additional harmonic oscillator (\ref{app01}). It is obvious that expression (\ref{app19})
reduces to the Green function of the harmonic oscillator for $\omega\to 0$:
\begin{equation}
\label{app19a}
\lim_{\omega\to 0} G_{ii}^{{\bf x}_0}(\tau,\tau')=\frac{1}{M\beta\Omega^2}\left(
\frac{\hbar\beta\Omega}{2}\,\frac{\cosh{\Omega(|\tau-\tau'|-\hbar\beta/2)}}
{\sinh{\hbar\beta\Omega/2}}-1\right)
\end{equation}
with $i\in\{x,y\}$. In this limit, the partition function (\ref{app15}) turns out to be the
usual one~\cite[Chap. 5]{PI} for such a harmonic oscillator
\begin{equation}
\label{app19aa}
\lim_{\omega\to 0} Z^{{\bf x}_0}=\frac{\hbar\beta\Omega/2}{\sinh{\hbar\beta\Omega/2}}.
\end{equation}
It is worth mentioning that with the last term in Green function (\ref{app19}) the 
classical harmonic fluctuation width 
\begin{equation}
\label{app19b}
G^{\rm cl}_{xx}=\left\langle x^2\right\rangle^{\rm cl}=\frac{1}{M\beta\Omega^2}
\end{equation}
is subtracted. This is the consequence of the
exclusion of the zero frequency mode of the Fourier path (\ref{app02}) in the generating
functional (\ref{app00}). The necessity to do this has already been discussed in 
Sect.~\ref{effrep}. The other terms in Eq.~(\ref{app19}) are those which we would have 
obtained {\it without} separation of the $x_0$-component. Thus these terms represent the
quantum mechanical Green function containing all quantum as well as thermal fluctuations. 
It is a nice property of all Green functions discussed in this paper that
\begin{equation}
\label{app19c}
G^{{\bf x}_0}_{xx}(\tau,\tau')=G^{\rm qm}_{xx}(\tau,\tau')-G^{\rm cl}_{xx}.
\end{equation}
Such a relation exists for all other Green functions appropriately, including momentum-position 
correlations which we consider subsequently.

The knowledge of relation (\ref{app18}) makes it quite easy to determine the algebraic
expression for the mixed $xy$-correlation functions. Rewriting Eq.~(\ref{app12b})
as
\begin{equation}
\label{app20}
G_{xy}^{{\bf x}_0}(\tau,\tau')=-G_{yx}^{{\bf x}_0}(\tau,\tau')
= \frac{i\omega}{M\beta}\,\frac{1}{\Omega_+^2-\Omega_-^2}\frac{\partial}{\partial \tau}
\left(\sum\limits_{m=-\infty}^\infty\,\frac{1}{\omega_m^2+\Omega_+^2}\,e^{-i\omega_m(\tau-\tau')}
+\sum\limits_{m=-\infty}^\infty\,\frac{1}{\omega_m^2+\Omega_-^2}\,e^{-i\omega_m(\tau-\tau')}
\right)
\end{equation}
and applying the derivative with respect to $\tau$ to relation (\ref{app18}), we obtain
the following expression for the mixed Green function:
\begin{eqnarray}
\label{app21}
G_{xy}^{{\bf x}_0}(\tau,\tau')&=&-G_{yx}^{{\bf x}_0}(\tau,\tau')\nonumber\\
&=&\frac{\hbar\omega}{2iM}\,\frac{1}{\Omega_+^2-\Omega_-^2}\left\{  
\Theta(\tau-\tau')[g_{\Omega_+}(\tau,\tau')-g_{\Omega_-}(\tau,\tau')]-
\Theta(\tau'-\tau)[g_{\Omega_+}(\tau',\tau)-g_{\Omega_-}(\tau',\tau)]
\right\},
\end{eqnarray}
where we have used the abbreviation
\begin{equation}
\label{app22}
g_{\Omega_\pm}(\tau,\tau')=\frac{\sinh{\Omega_\pm(\tau-\tau'-\hbar\beta/2)}}
{\sinh{\hbar\beta\Omega_\pm/2}},\qquad \tau,\tau'\in(0,\hbar\beta).
\end{equation}
Note that classically $\left\langle xy\right\rangle^{\rm cl}=0$ such that Eq.~(\ref{app19c}) 
reduces to 
\begin{equation}
\label{app22a}
G_{xy}^{{\bf x}_0}(\tau,\tau')=G_{xy}^{\rm qm}(\tau,\tau').
\end{equation}
The Heaviside function in Eq.~(\ref{app21}) is defined symmetrically:
\begin{equation}
\label{app22b}
\Theta(\tau-\tau')=\left\{\begin{array}{cc} 1 & \quad\tau>\tau',\\ 1/2 & \quad\tau=\tau', \\
0 & \quad\tau<\tau'. \end{array}\right.
\end{equation}
In the quantum mechanical limit of zero-temperature ($\beta\to \infty$), the Green function
(\ref{app19}) simplifies to 
\begin{equation}
\label{app23a}
\lim_{\beta\to\infty} G_{xx}^{{\bf x}_0}(\tau,\tau')=
\lim_{\beta\to\infty} G_{yy}^{{\bf x}_0}(\tau,\tau')=\frac{\hbar}{2M}
\left(\frac{1}{\Omega_+}\,\frac{\Omega_+^2-\Omega^2}{\Omega_+^2-\Omega_-^2}\,
e^{-\Omega_+|\tau-\tau'|}-\frac{1}{\Omega_-}\,\frac{\Omega_-^2-\Omega^2}{\Omega_+^2-\Omega_-^2}\,
e^{-\Omega_-|\tau-\tau'|}\right),
\end{equation}
while in Eq.~(\ref{app21}) only $g_{\Omega_\pm}(\tau,\tau')$ changes:
\begin{equation}
\label{app23b}
\lim_{\beta\to\infty} g_{\Omega_\pm}(\tau,\tau')=-e^{-\Omega_\pm(\tau-\tau')}.
\end{equation}
\section{Properties of Green Functions}
\label{appprop}
In this section we list properties of the Green functions (\ref{app19}) and (\ref{app21})
which are important for the forthcoming consideration of the generating functional with
sources coupling linearily to position or momentum in Appendix~\ref{appgenrp}. For all
relations we suppose that $0\le\tau,\tau'\le\hbar\beta$.
\subsection{General Properties}
A first observation is the temporal translational invariance of the Green functions:
\begin{equation}
\label{prop00}
G_{ij}^{{\bf x}_0}(\tau,\tau')=G_{ij}^{{\bf x}_0}(\tau-\tau'),
\end{equation}
where each of the indices $i,j$ stands for $x$ or $y$, respectively. For equal times we
find
\begin{equation}
\label{prop01}
G_{ij}^{{\bf x}_0}(\tau,\tau)=\frac{1}{M\beta}
\left(\frac{\hbar\beta}{2\Omega_+}\,\frac{\Omega_+^2-\Omega^2}{\Omega_+^2-\Omega_-^2} 
\coth{\hbar\beta\Omega_+/2}-\frac{\hbar\beta}{2\Omega_-}\,
\frac{\Omega_-^2-\Omega^2}{\Omega_+^2-\Omega_-^2} \coth{\hbar\beta\Omega_-/2}-\frac{1}{\Omega^2}
\right)\times
\left\{\begin{array}{cc}
1 & \quad i=j,\\
0 & \quad i\neq j.
\end{array}\right.
\end{equation}
Moreover we read off the following symmetries from the expressions (\ref{app19}) and
(\ref{app21}):
\begin{equation}
\label{prop02}
G_{ij}^{{\bf x}_0}(\tau,\tau')=G_{ij}^{{\bf x}_0}(\tau',\tau)\times\left\{\begin{array}{cc}
1 & \quad i=j,\\
-1 & \quad i\neq j.
\end{array}\right.
\end{equation}
Otherwise,
\begin{equation}
\label{prop03}
G_{ij}^{{\bf x}_0}(\tau,\tau')=G_{ji}^{{\bf x}_0}(\tau',\tau).
\end{equation}
Throughout the paper we always use periodic paths. Hence it is obvious that all
Green functions are periodic, too:
\begin{equation}
\label{prop04}
G_{ij}^{{\bf x}_0}(0,\tau')=G_{ij}^{{\bf x}_0}(\hbar\beta,\tau'),\qquad 
G_{ij}^{{\bf x}_0}(\tau,0)=G_{ij}^{{\bf x}_0}(\tau,\hbar\beta).
\end{equation}
\subsection{Derivatives of Green Functions}
We now proceed with derivatives of the Green functions (\ref{app19}) and
(\ref{app21}), since these are essential for the derivation of the generating 
functional of position and momentum dependent correlations in
the forthcoming Appendix~\ref{appgenrp}. 

Before considering the concrete expressions we introduce a new symbol indicating
uniquely to which argument the derivative is applied. A dot on the left-hand side
means to perform the derivative with respect to the first argument and the dot
on the right-hand side indicates that to differentiate with respect to the other
argument. Having a dot on both sides the Green function is derived with respect
to both arguments:
\begin{equation}
\label{der00}
\dg_{ij}^{{\bf x}_0}(\tau,\tau')=
\frac{\partial G_{ij}^{{\bf x}_0}(\tau,\tau')}{\partial \tau},\quad
\gd_{ij}^{{\bf x}_0}(\tau,\tau')=
\frac{\partial G_{ij}^{{\bf x}_0}(\tau,\tau')}{\partial \tau'},\quad
\dgd_{ij}^{{\bf x}_0}(\tau,\tau')=
\frac{\partial^2 G_{ij}^{{\bf x}_0}(\tau,\tau')}{\partial \tau \partial \tau'}.
\end{equation}
Applying such derivatives to the Green functions (\ref{app19}), we obtain ($i\in \{x,y\}$):
\begin{equation}
\label{der01}
\dg_{ii}^{{\bf x}_0}(\tau,\tau')=\frac{\hbar}{2M}\frac{1}{\Omega_+^2-\Omega_-^2}
\left[\Theta(\tau-\tau')g(\tau,\tau') 
-\Theta(\tau'-\tau)g(\tau',\tau)\right]=-\gd_{ii}^{{\bf x}_0}(\tau,\tau')
\end{equation}
with 
\begin{equation}
\label{der02}
g(\tau,\tau')=(\Omega_+^2-\Omega^2)g_{\Omega_+}(\tau,\tau')
-(\Omega_-^2-\Omega^2)g_{\Omega_-}(\tau,\tau'),
\end{equation}
where $g_\pm(\tau,\tau)$ was defined in Eq.~(\ref{app22}).
Performing the derivatives to both arguments leads to the expression
\begin{equation}
\label{der03}
\dgd_{ii}^{{\bf x}_0}(\tau,\tau')=\tdgd_{ii}^{{\bf x}_0}(\tau,\tau')+
\frac{\hbar}{M}\delta(\tau-\tau'),
\end{equation}
where we have introduced the partial function 
\begin{equation}
\label{der04}
\tdgd_{ii}^{{\bf x}_0}(\tau,\tau')=-\frac{\hbar}{2M}\left[\Omega_+
\frac{\Omega_+^2-\Omega^2}{\Omega_+^2-\Omega_-^2}
\frac{\sinh{\Omega_+(|\tau-\tau'|-\hbar\beta/2)}}{\sinh{\hbar\beta\Omega_+/2}}-
\Omega_-
\frac{\Omega_-^2-\Omega^2}{\Omega_+^2-\Omega_-^2}
\frac{\sinh{\Omega_-(|\tau-\tau'|-\hbar\beta/2)}}{\sinh{\hbar\beta\Omega_-/2}} \right]
\end{equation}
which is finite for equal times.

Applying derivatives with respect to the first respective second argument to the mixed
correlation function (\ref{app21}), we find:
\begin{equation}
\label{der05}
\dg_{xy}^{{\bf x}_0}(\tau,\tau')=\frac{\hbar\omega}{2iM}\frac{1}{\Omega_+^2-\Omega_-^2}
\left[\Omega_+\frac{\cosh{\Omega_+(|\tau-\tau'|-\hbar\beta/2)}}{\sinh{\hbar\beta\Omega_+/2}}
-\Omega_-\frac{\cosh{\Omega_-(|\tau-\tau'|-\hbar\beta/2)}}{\sinh{\hbar\beta\Omega_-/2}}\right]
=-\gd_{xy}^{{\bf x}_0}(\tau,\tau')
\end{equation}
and
\begin{equation}
\label{der05b}
\dg_{yx}^{{\bf x}_0}(\tau,\tau')=-\dg_{xy}^{{\bf x}_0}(\tau,\tau').
\end{equation}
Differentiating each argument of the mixed Green function results in
\begin{equation}
\label{der06}
\dgd_{xy}^{{\bf x}_0}(\tau,\tau')=\frac{i\hbar\omega}{2M}\frac{1}{\Omega_+^2-\Omega_-^2}
\left[\Theta(\tau-\tau')h(\tau,\tau')-\Theta(\tau'-\tau)h(\tau',\tau) \right]=
-\dgd_{yx}^{{\bf x}_0}(\tau,\tau')
\end{equation}
with
\begin{equation}
\label{der07}
h(\tau,\tau')=\Omega_+^2g_{\Omega_+}(\tau,\tau')-\Omega_-^2g_{\Omega_-}(\tau,\tau').
\end{equation}
An additional property we read off from Eqs.~(\ref{der01}) and (\ref{der05}) is
($i,j\in \{x,y\}$):
\begin{eqnarray}
\label{der08a}
\dg_{ij}^{{\bf x}_0}(\tau,\tau')=\dg_{ij}^{{\bf x}_0}(\tau',\tau)\times
\left\{\begin{array}{cc}
-1 &\quad i=j,\\
1 & \quad i\neq j,
\end{array}\right.\\
\label{der08b}
\gd_{ij}^{{\bf x}_0}(\tau,\tau')=\gd_{ij}^{{\bf x}_0}(\tau',\tau)\times
\left\{\begin{array}{cc}
-1 &\quad i=j,\\
1 & \quad i\neq j.
\end{array}\right.
\end{eqnarray}
The double-sided derivatives (\ref{der03}), (\ref{der04}), and (\ref{der06}) imply
\begin{equation}
\label{der09}
\dgd_{ij}^{{\bf x}_0}(\tau,\tau')=\dgd_{ij}^{{\bf x}_0}(\tau',\tau)\times
\left\{\begin{array}{cc}
1 &\quad i=j,\\
-1 & \quad i\neq j.
\end{array}\right.
\end{equation}
The derivatives (\ref{der01}), (\ref{der04}), (\ref{der05}), and (\ref{der06}) 
are periodic:
\begin{eqnarray}
\label{der10a}
&&\dg_{ij}^{{\bf x}_0}(\tau,0)=\dg_{ij}^{{\bf x}_0}(\tau,\hbar\beta),\quad 
\dg_{ij}^{{\bf x}_0}(0,\tau')=\dg_{ij}^{{\bf x}_0}(\hbar\beta,\tau'),\\
\label{der10b}
&&\gd_{ij}^{{\bf x}_0}(\tau,0)=\gd_{ij}^{{\bf x}_0}(\tau,\hbar\beta),\quad 
\gd_{ij}^{{\bf x}_0}(0,\tau')=\gd_{ij}^{{\bf x}_0}(\hbar\beta,\tau'),\\
\label{der10c}
&&\tdgd_{ii}^{{\bf x}_0}(\tau,0)=\tdgd_{ii}^{{\bf x}_0}(\tau,\hbar\beta),\quad 
\tdgd_{ii}^{{\bf x}_0}(0,\tau')=\tdgd_{ii}^{{\bf x}_0}(\hbar\beta,\tau'),\\
\label{der10d}
&&\dgd_{ij}^{{\bf x}_0}(\tau,0)=\dgd_{ij}^{{\bf x}_0}(\tau,\hbar\beta),\quad 
\dgd_{ij}^{{\bf x}_0}(0,\tau')=\dgd_{ij}^{{\bf x}_0}(\hbar\beta,\tau'),\qquad (i\neq j).
\end{eqnarray}
\section{Generating Functional for Position- and Momentum-Dependent Correlation Functions}
\label{appgenrp}
With the discussion of the generating functional for position-dependent correlation
functions and, in particular, the Green functions in Appendix~\ref{appgen} and their
properties in Appendix~\ref{appprop}, we have layed the foundation to derive the 
generating functional for correlation functions depending on both, position and momentum.
Following the framework presented in an earlier work~\cite{correlation}, such a functional
involving sources coupled to the momentum can always be reduced to one containing
position-coupled sources only. 

We start from the three-dimensional effective classical representation for the generating 
functional
\begin{equation}
\label{xp00}
Z_{\bf \Omega}[{\bf j},{\bf v}]=
\int\frac{d^3x_0d^3p_0}{(2\pi\hbar)^3}Z^{{\bf p}_0,{\bf x}_0}_{\bf \Omega}[{\bf j},{\bf v}]
\end{equation}
with zero frequency components ${\bf x}_0=(x_0,y_0,z_0)={\rm const.}$ and 
${\bf p}_0=({p_x}_0,{p_y}_0,{p_y}_0)={\rm const.}$ of the Fourier path separated. 
The reduced functional is
\begin{equation}
\label{xp01}
Z^{{\bf p}_0,{\bf x}_0}_{\bf \Omega}[{\bf j},{\bf v}]=
(2\pi\hbar)^3\oint {\cal D'}^3x{\cal D}^3p\,\delta({\bf x}_0-\overline{{\bf x}(\tau)})
\delta({\bf p}_0-\overline{{\bf p}(\tau)})\,
\exp\left\{-\frac{1}{\hbar}{\cal A}^{{\bf p}_0,{\bf x}_0}_{\bf \Omega}
[{\bf p},{\bf x};{\bf j},{\bf v}] \right\},
\end{equation}  
where the path integral measure is that defined in Eq.~(\ref{ham03}). 
Extending the action (\ref{ham02}) by source terms, considering a more general Hamilton
function than (\ref{ham16}), and introducing an additional harmonic oscillator in $z$-direction, 
the action functional in Eq.~(\ref{xp01}) shall read
\begin{eqnarray}
\label{xp02}
{\cal A}^{{\bf p}_0,{\bf x}_0}_{\bf \Omega}[{\bf p},{\bf x};{\bf j},{\bf v}]=
\int_0^{\hbar\beta}d\tau\,\Big\{&&-i\tp(\tau)\cdot \dot{\bf x}(\tau)+
\frac{1}{2M}\tp^2(\tau)-\frac{1}{2}\osa l_z(\tp,\tr) 
+\frac{1}{8}M\osb^2\left[ 
\tx^2(\tau)+\ty^2(\tau)\right]+\frac{1}{2}M\op^2\tz^2(\tau)\nonumber\\
&&+{\bf j}(\tau)\cdot\tr(\tau)+{\bf v}(\tau)\cdot\tp(\tau) 
\Big\}
\end{eqnarray}
with shifted positions and momenta 
\begin{equation}
\label{xp03}
\tr={\bf x}(\tau)-{\bf x}_0,\qquad \tp={\bf p}(\tau)-{\bf p}_0.
\end{equation} 
The orbital angular momentum $l_z({\bf p},{\bf x})$ is defined in Eq.~(\ref{ham18}) and is 
used in Eq.~(\ref{xp02}) with the shifted phase space coordinates (\ref{xp03}). We
have introduced three different frequencies in (\ref{xp02}), ${\bf \Omega}=(\osa,\osb,\op)$,
where the first both components are used in regard to the oscillations in the plane 
perpendicular to the direction of the magnetic field which shall be considered here to point
into $z$-direction. The last component, $\op$, is the frequency of a trial oscillator 
parallel to the field lines. 

Due to the periodicity of the paths, we suppose that the sources might also be
periodic:
\begin{equation}
\label{xp03b}
{\bf j}(0)={\bf j}(\hbar\beta),\qquad {\bf v}(0)={\bf v}(\hbar\beta).
\end{equation}
Since we want to simplify expression (\ref{xp01}) such that we can use the results 
obtained in Appendix~\ref{appgen}, the momentum path integral is solved in the 
following. In a first step we reexpress the momentum $\delta$-function in (\ref{xp01}) by
\begin{equation}
\label{xp04}
\delta({\bf p}_0-\overline{{\bf p}(\tau)})=\int\frac{d^3\xi}{(2\pi\hbar)^3}\,
\exp\left\{-\frac{1}{\hbar}\int_0^{\hbar\beta} d\tau \,{\bf v}_0\cdot[{\bf p}(\tau)-{\bf p}_0]
\right\},
\end{equation}
where 
\begin{equation}
\label{xp05}
{\bf v}_0(\bold{\xi})=\frac{i}{\hbar\beta}\bold{\xi}
\end{equation}
is an additional current which is coupled to the momentum and is constant in time. Defining
the sum of all sources coupled to the momentum by 
\begin{equation}
\label{xp06}
{\bf V}(\bold{\xi},\tau)={\bf v}(\tau)+{\bf v}_0(\bold{\xi}),
\end{equation}
the functional (\ref{xp01}) can be written as
\begin{eqnarray}
\label{xp07}
Z^{{\bf p}_0,{\bf x}_0}_{\bf \Omega}[{\bf j},{\bf v}]&=&
\int d^3\xi\oint {\cal D'}^3x{\cal D}^3p\,\delta({\bf x}_0-\overline{{\bf x}(\tau)})
\exp\Bigg\{-\frac{1}{\hbar}\int_0^{\hbar\beta}d\tau\,\Big[-i{\bf p}(\tau)
\cdot\dot{\bf x}(\tau) 
+\frac{{\bf p}^2(\tau)}{2M}-\frac{1}{2}\osa l_z({\bf p}(\tau),\tr(\tau))\nonumber\\
&&+\frac{1}{8}M\osb^2\left\{\tx^2(\tau)+\ty^2(\tau)\right\}+\frac{1}{2}M\op^2\tz^2(\tau)
+{\bf j}(\tau)\cdot\tr(\tau)+{\bf V}(\bold{\xi},\tau)\cdot{\bf p}(\tau)
\Big] \Bigg\},
\end{eqnarray}
where we have used the translation invariance $\tp\to {\bf p}$ of the path integral. 
To solve the momentum path integral, it is useful to express it in its discretized form. 
Performing quadratic completions such that the momentum path integral separates into
an infinite product of simple Gaussian integrals which are easily calculated, the remaining
functional is reduced to the configuration space path integral
\begin{eqnarray}
\label{xp08}
Z^{{\bf p}_0,{\bf x}_0}_{\bf \Omega}[{\bf j},{\bf v}]=
\int d^3\xi\,\exp\left[\frac{M}{2\hbar}\int_0^{\hbar\beta}d\tau\,{\bf V}^2(\bold{\xi},\tau) 
\right]
\oint{\cal D}^3x\,\delta({\bf x}_0-\overline{{\bf x}(\tau)})
\exp\left\{-\frac{1}{\hbar}{\cal A}^{{\bf p}_0,{\bf x}_0}_{\bf \Omega}
[{\bf x};{\bf j},{\bf V}]\right\}
\end{eqnarray}
with the measure (\ref{ham09}) for $D=3$. The action functional is
\begin{eqnarray}
\label{xp09}
{\cal A}^{{\bf p}_0,{\bf x}_0}_{\bf \Omega}[{\bf x};{\bf j},{\bf V}]&=&
\int_0^{\hbar\beta}d\tau\,\Bigg[\frac{M}{2}\dot{\bf x}^2(\tau)  
+\frac{1}{2}iM\osa\left\{\dot{x}(\tau)\ty(\tau)-\dot{y}(\tau)\tx(\tau)\right\}
+\frac{1}{8}M\left(\osb^2-\osa^2\right)\left\{\tx^2(\tau)+\ty^2(\tau)\right\}\nonumber\\
&&\hspace*{-3pt}
+\frac{1}{2}M\op^2\tz^2(\tau)+\tx(\tau)\left[j_x(\tau)+\frac{1}{2}M\osa V_y(\bold{\xi},\tau)
\right]\nonumber\\
&&\hspace*{-3pt}+\ty(\tau)\left[j_y(\tau)-\frac{1}{2}M\osa V_x(\bold{\xi},\tau)\right]+\tz(\tau)j_z(\tau)
\Bigg]
-\frac{iM}{\hbar}\int_0^{\hbar\beta}d\tau\,\dot{\bf x}(\tau)\cdot
{\bf V}(\bold{\xi},\tau),
\end{eqnarray}
where the last term simplifies by the following consideration. A partial integration of this 
term yields
\begin{equation}
\label{xp10}
\int_0^{\hbar\beta}d\tau\,\dot{\bf x}(\tau)\cdot{\bf V}(\bold{\xi},\tau)
=-\int_0^{\hbar\beta}d\tau\,({\bf x}(\tau)-{\bf x}_0)\cdot\dot{\bf V}(\bold{\xi},\tau).
\end{equation}
The surface term vanishes as a consequence of the periodicity of the path and the source.
This periodicity is also the reason why we could shift ${\bf x}(\tau)$ by the constant
${\bf x}_0$ on the right-hand side of Eq.~(\ref{xp10}). Obviously, the importance of this
expression lies in the coupling of the time derivative of ${\bf V}(\bold{\xi},\tau)$ to
the path ${\bf x}(\tau)$. Thus, $\dot{\bf V}(\bold{\xi},\tau)$ can be handled like a 
${\bf j}(\tau)$-current~\cite{correlation} and the action (\ref{xp09}) can be written as
\begin{equation}
\label{xp11}
{\cal A}^{{\bf p}_0,{\bf x}_0}_{\bf \Omega}[{\bf x};{\bf j},{\bf V}]=
{\cal A}^{{\bf p}_0,{\bf x}_0}_{\bf \Omega}[{\bf x};{\bf J},0]=
{\cal A}^{{\bf p}_0,{\bf x}_0}_{\bf \Omega}[{\bf x};0,0]-\frac{1}{\hbar}
\int_0^{\hbar\beta}d\tau\,\tr(\tau)\cdot{\bf J}(\bold{\xi},\tau)
\end{equation}
with the new current vector ${\bf J}(\bold{\xi},\tau)$ which has the components
\begin{eqnarray}
\label{xp12}
J_x(\bold{\xi},\tau)&=&j_x(\tau)+\frac{1}{2}M\osa V_y(\bold{\xi},\tau)-
iM\dot{V}_x(\bold{\xi},\tau),\nonumber\\ 
J_y(\bold{\xi},\tau)&=&j_y(\tau)-\frac{1}{2}M\osa V_x(\bold{\xi},\tau)-
iM\dot{V}_y(\bold{\xi},\tau),\\
J_z(\bold{\xi},\tau)&=&j_z(\tau)-\frac{1}{2}M\op V_z(\bold{\xi},\tau)\nonumber.
\end{eqnarray}
and couples to the path ${\bf x}(\tau)$ only. 
With the expression (\ref{xp08}) for the 
generating functional and the action (\ref{xp11}), we have derived a representation
similar to Eq.~(\ref{app00}) with the action (\ref{app01}), extended by an additional
oscillator in $z$-direction. We identify
\begin{equation}
\label{xp13}
\omega\equiv \osa,\quad \Omega\equiv \frac{1}{4}\left(\osb^2-\osa^2\right),
\quad j_x\equiv J_x,\quad
j_y\equiv J_y.
\end{equation}
Thus the auxiliary frequencies $\Omega_{\pm}$ (\ref{app14}) become
\begin{equation}
\label{xp13b}
\Omega_{\pm}(\osa,\osb)=\frac{1}{2}\,|\osa\pm\osb|.
\end{equation}
Inserting the substitutions (\ref{xp13})
into the solution (\ref{app07}) for the generating functional in two dimensions 
and performing the usual calculation for a harmonic oscillator with 
external source~\cite[Chaps. 3,5]{PI} in $z$-direction, we obtain an intermediate result for  
the generating functional in three dimensions (\ref{xp01}):
\begin{equation}
\label{xp14}
Z^{{\bf p}_0,{\bf x}_0}_{\bf \Omega}[{\bf j},{\bf v}]=
\lambda_{\rm th}^{-3}\,Z^{{\bf p}_0,{\bf x}_0}_{\bf \Omega}
\int d^3\xi\,\exp\left\{\frac{M}{2\hbar}\int_0^{\hbar\beta}d\tau\,{\bf V}^2(\bold{\xi},\tau) 
\right\}\,\exp\left\{\frac{1}{2\hbar^2}\int_0^{\hbar\beta}d\tau\int_0^{\hbar\beta}d\tau'\,
{\bf J}(\bold{\xi},\tau)\,{\bf G}^{{\bf x}_0}(\tau,\tau')\,{\bf J}(\bold{\xi},\tau')\right\}.
\end{equation}
The partition function follows from Eqs.~(\ref{app15}) and (\ref{app19aa})
\begin{equation}
\label{xp15}
Z^{{\bf p}_0,{\bf x}_0}_{\bf \Omega}=Z^{{\bf p}_0,{\bf x}_0}_{\bf \Omega}[0,0]=
\frac{\hbar\beta\Omega_+/2}{\sinh{\hbar\beta\Omega_+/2}}\,
\frac{\hbar\beta\Omega_-/2}{\sinh{\hbar\beta\Omega_-/2}}\,
\frac{\hbar\beta\op/2}{\sinh{\hbar\beta\op/2}}
\end{equation}
and ${\bf G}^{{\bf x}_0}(\tau,\tau')$ is the $3\times 3$-matrix of Green functions
\begin{equation}
\label{xp16}
{\bf G}^{{\bf x}_0}(\tau,\tau')=\left(\begin{array}{ccc}
G_{xx}^{{\bf x}_0}(\tau,\tau') & G_{xy}^{{\bf x}_0}(\tau,\tau') & 0\\
G_{yx}^{{\bf x}_0}(\tau,\tau') & G_{yy}^{{\bf x}_0}(\tau,\tau') & 0\\
0 & 0 & G_{zz}^{{\bf x}_0}(\tau,\tau')\end{array}\right).
\end{equation}
Except $G_{zz}^{{\bf x}_0}(\tau,\tau')$, the Green functions are given by the expressions
in Eqs.~(\ref{app19}) and (\ref{app21}) with frequencies (\ref{xp13b}). 
The Green function of the pure harmonic oscillator in $z$-direction
\begin{equation}
\label{xp17}
G_{zz}^{{\bf x}_0}(\tau,\tau')=\frac{1}{M\beta\op^2}\left(
\frac{\hbar\beta\op}{2}\,\frac{\cosh{\op(|\tau-\tau'|-\hbar\beta/2)}}
{\sinh{\hbar\beta\op/2}}-1\right)
\end{equation}
follows directly from the limit (\ref{app19a}).
Since the current ${\bf J}$ (\ref{xp12}) still depends on time derivatives of ${\bf V}$,
we have to perform some partial integrations in the functional (\ref{xp14}). This is a 
very extensive but straightforward work and thus we only present an instructive example.
For that we apply the properties and the time derivatives of the Green functions 
which we presented in Appendix~\ref{appprop}.
Consider the integral 
\begin{equation}
\label{xp18}
I=-\frac{M^2}{2\hbar^2}\int_0^{\hbar\beta}d\tau\int_0^{\hbar\beta}d\tau'\,\dot{V}_i
(\bold{\xi},\tau)\,G_{ii}^{{\bf x}_0}(\tau,\tau')\,\dot{V}_i(\bold{\xi},\tau')
\end{equation}
occuring in the second exponential of Eq.~(\ref{xp14}) with $i\in\{x,y,z\}$. 
A partial integration in the $\tau'$-integral leads to
\begin{eqnarray}
\label{xp18a}
I&=&-\frac{M^2}{2\hbar^2}\int_0^{\hbar\beta}d\tau\,\dot{V}_i(\bold{\xi},\tau)
\left(G_{ii}^{{\bf x}_0}(\tau,\tau')\,V_i(\bold{\xi},\tau')\Big|_{\tau'=0}^{\tau'=\hbar\beta}
-\int_0^{\hbar\beta}d\tau'\,\frac{\partial G_{ii}^{{\bf x}_0}(\tau,\tau')}{\partial \tau'}
\,V_i(\bold{\xi},\tau')\right)\nonumber\\
&=&\frac{M^2}{2\hbar^2}\int_0^{\hbar\beta}d\tau\int_0^{\hbar\beta}d\tau'\,
\dot{V}_i(\bold{\xi},\tau)\,\gd_{ii}^{{\bf x}_0}(\tau,\tau')\,V_i(\bold{\xi},\tau').
\end{eqnarray}
The surface term in the first line vanishes as a consequence of the periodicity
of the current (\ref{xp03b}) and the Green function (\ref{prop04}). A second partial
integration, now in the $\tau$-integral, results in
\begin{eqnarray}
\label{xp18b}
I&=&-\frac{M^2}{2\hbar^2}\int_0^{\hbar\beta}d\tau\int_0^{\hbar\beta}d\tau'\,
V_i(\bold{\xi},\tau)\,\dgd_{ii}^{{\bf x}_0}(\tau,\tau')\,V_i(\bold{\xi},\tau')\nonumber\\
&=&-\frac{M^2}{2\hbar^2}\int_0^{\hbar\beta}d\tau\int_0^{\hbar\beta}d\tau'\,
V_i(\bold{\xi},\tau)\,\tdgd_{ii}^{{\bf x}_0}(\tau,\tau')\,V_i(\bold{\xi},\tau')-
\frac{M}{2\hbar}\int_0^{\hbar}d\tau\,V_i^2(\bold{\xi},\tau).
\end{eqnarray}
Here we have applied the periodicity property of the right-hand derivative of the Green function 
(\ref{der10b}), leading to a vanishing surface term in this case, too. In the second line, we 
have used the decomposition (\ref{der03}) of the double-sided differentiated Green function.
Note that the last term just cancels the appropriate term in the first exponential of 
the right-hand side of Eq.~(\ref{xp14}). Eventually, after performing all such partial 
integrations, we reexpress Eq.~(\ref{xp14}) by
\begin{equation}
\label{xp19}
Z^{{\bf p}_0,{\bf x}_0}_{\bf \Omega}[{\bf j},{\bf v}]=
\lambda_{\rm th}^{-3}\,Z^{{\bf p}_0,{\bf x}_0}_{\bf \Omega}
\int d^3\xi\,\exp\left\{\frac{1}{2\hbar^2}\int_0^{\hbar\beta}d\tau\int_0^{\hbar\beta}d\tau'\,
\tilde{\bf s}(\bold{\xi},\tau)\,{\bf H}^{{\bf x}_0}(\tau,\tau')\,\tilde{\bf s}(\bold{\xi},\tau')  
\right\}
\end{equation}
with six-dimensional sources
\begin{equation}
\label{xp20}
\tilde{\bf s}(\bold{\xi},\tau)=\left(\,{\bf j}(\tau),{\bf V}(\bold{\xi},\tau)\,\right).
\end{equation}
and the $6\times 6$-matrix ${\bf H}^{{\bf x}_0}(\tau,\tau')$ which has no significance
as long as we have not done the $\bold{\xi}$-integration. We explicitly insert 
the decomposition (\ref{xp06}) into expression (\ref{xp20}) of the source vector $\tilde{\bf s}$.
Since ${\bf v}_0(\bold{\xi})$ from Eq.~(\ref{xp05}) is constant in time, 
some temporal integrals in the exponential of Eq.~(\ref{xp19}) can be calculated and 
we obtain
\begin{eqnarray}
\label{xp21}
Z^{{\bf p}_0,{\bf x}_0}_{\bf \Omega}[{\bf j},{\bf v}]&=&
\lambda_{\rm th}^{-3}\,Z^{{\bf p}_0,{\bf x}_0}_{\bf \Omega}
\exp\left\{\frac{1}{2\hbar^2}\int_0^{\hbar\beta}d\tau\int_0^{\hbar\beta}d\tau'\,
{\bf s}(\tau)\,{\bf H}^{{\bf x}_0}(\tau,\tau')\,{\bf s}(\tau')\right\}\nonumber\\
&&\times\int d^3\xi\,
\exp\left\{-\frac{M}{2\hbar^2\beta}\bold{\xi}^2+i\frac{M}{\hbar^2\beta}\bold{\xi}\cdot
\int_0^{\hbar\beta}d\tau\,{\bf v}(\tau) \right\}
\end{eqnarray}
with the new $6$-vector
\begin{equation}
\label{xp22}
{\bf s}(\tau)=(\,{\bf j}(\tau),{\bf v}(\tau)\,)
\end{equation}
consisting of the original sources ${\bf j}$ and ${\bf v}$ only. The Gaussian $\xi$-integral in
Eq.~(\ref{xp21}) can be easily solved and the terms appearing from quadratic completion
modify the above matrix ${\bf H}^{{\bf x}_0}(\tau,\tau')$. The final result for the
generating functional of all position and momentum dependent correlations is given by
\begin{equation}
\label{xp23}
Z^{{\bf p}_0,{\bf x}_0}_{\bf \Omega}[{\bf j},{\bf v}]=Z^{{\bf p}_0,{\bf x}_0}_{\bf \Omega}
\exp\left\{\frac{1}{2\hbar^2}\int_0^{\hbar\beta}d\tau\int_0^{\hbar\beta}d\tau'\,
{\bf s}(\tau)\,{\bf G}^{{\bf p}_0,{{\bf x}_0}}(\tau,\tau')\,{\bf s}(\tau')\right\}.
\end{equation}
The complete $6\times 6$-matrix ${\bf G}^{{\bf p}_0,{{\bf x}_0}}(\tau,\tau')$ 
contains all possible Green functions describing
position-position, position-momentum, and momentum-momentum correlations. As a consequence
of separating the fluctuations into those perpendicular and parallel to the direction
of the magnetic field, all correlations between $x,y$ on the one and $z$ on the other hand
vanish as well as those for the appropriate momenta. The symmetries for the Green functions
and their derivatives were investigated in detail in Appendix~\ref{appprop} and lead to
a further reduction of the number of significant matrix elements. It turns out that
only 9 elements are independent of each other. Therefore we can write the matrix
\begin{equation}
\label{xp24}
{\bf G}^{{\bf x}_0,{{\bf p}_0}}(\tau,\tau')=\left(\begin{array}{cccccc}
G_{xx}^{{\bf p}_0,{\bf x}_0}(\tau,\tau') & G_{xy}^{{\bf p}_0,{\bf x}_0}(\tau,\tau') & 0 & 
G_{xp_x}^{{\bf p}_0,{\bf x}_0}(\tau,\tau') & G_{xp_y}^{{\bf p}_0,{\bf x}_0}(\tau,\tau') & 0 \\
G_{xy}^{{\bf p}_0,{\bf x}_0}(\tau',\tau) & G_{xx}^{{\bf p}_0,{\bf x}_0}(\tau,\tau') & 0 & 
-G_{xp_y}^{{\bf p}_0,{\bf x}_0}(\tau,\tau') & G_{xp_x}^{{\bf p}_0,{\bf x}_0}(\tau,\tau') & 0 \\
0 & 0 & G_{zz}^{{\bf p}_0,{\bf x}_0}(\tau,\tau') & 0 & 0 & 
G_{zp_z}^{{\bf p}_0,{\bf x}_0}(\tau,\tau')\\
G_{xp_x}^{{\bf p}_0,{\bf x}_0}(\tau',\tau) & -G_{xp_y}^{{\bf p}_0,{\bf x}_0}(\tau',\tau) & 0 & 
G_{p_xp_x}^{{\bf p}_0,{\bf x}_0}(\tau,\tau') & G_{p_xp_y}^{{\bf p}_0,{\bf x}_0}(\tau,\tau') & 0 \\
G_{xp_y}^{{\bf p}_0,{\bf x}_0}(\tau',\tau) & G_{xp_x}^{{\bf p}_0,{\bf x}_0}(\tau',\tau) & 0 & 
G_{p_xp_y}^{{\bf p}_0,{\bf x}_0}(\tau',\tau) & G_{p_xp_x}^{{\bf p}_0,{\bf x}_0}(\tau,\tau') & 0 \\
0 & 0 & G_{zp_z}^{{\bf p}_0,{\bf x}_0}(\tau',\tau) & 0 & 0 & 
G_{p_zp_z}^{{\bf p}_0,{\bf x}_0}(\tau,\tau')
\end{array}\right).
\end{equation}
The matrix decays into four $3\times 3$-blocks,
each of the which describing another type of correlation: the upper left
position-position, the upper right position-momentum (as well as the lower left one), 
and the lower right momentum-momentum correlations. The different elements of the matrix are
\begin{eqnarray}
\label{xp25a}
G_{xx}^{{\bf p}_0,{\bf x}_0}(\tau,\tau')&=&\meanrp{\tx(\tau)\tx(\tau')}=
G_{xx}^{{\bf x}_0}(\tau,\tau'),\\
\label{xp25b}
G_{xy}^{{\bf p}_0,{\bf x}_0}(\tau,\tau')&=&\meanrp{\tx(\tau)\ty(\tau')}=
G_{xy}^{{\bf x}_0}(\tau,\tau'),\\
\label{xp25c}
G_{zz}^{{\bf p}_0,{\bf x}_0}(\tau,\tau')&=&\meanrp{\tz(\tau)\tz(\tau')}=
G_{zz}^{{\bf x}_0}(\tau,\tau'),\\ 
\label{xp25d}
G_{xp_x}^{{\bf p}_0,{\bf x}_0}(\tau,\tau')&=&\meanrp{\tx(\tau)\tpx(\tau')}=
iM\gd_{xx}^{{\bf x}_0}(\tau,\tau')-\frac{1}{2}M\osa G_{xy}^{{\bf x}_0}(\tau,\tau'),\\
\label{xp25e}
G_{xp_y}^{{\bf p}_0,{\bf x}_0}(\tau,\tau')&=&\meanrp{\tx(\tau)\tpy(\tau')}=
iM\gd_{xy}^{{\bf x}_0}(\tau,\tau')+\frac{1}{2}M\osa G_{xx}^{{\bf x}_0}(\tau,\tau'),\\
\label{xp25f}
G_{zp_z}^{{\bf p}_0,{\bf x}_0}(\tau,\tau')&=&\meanrp{\tz(\tau)\tpz(\tau')}=
iM\gd_{zz}^{{\bf x}_0}(\tau,\tau'),\\
\label{xp25g}
G_{p_xp_x}^{{\bf p}_0,{\bf x}_0}(\tau,\tau')&=&\meanrp{\tpx(\tau)\tpx(\tau')}=
-M^2\tdgd_{xx}^{{\bf x}_0}(\tau,\tau')-iM^2\osa\dg_{xy}^{{\bf x}_0}(\tau,\tau')+\frac{1}{4}
M^2\osa^2G_{xx}^{{\bf x}_0}(\tau,\tau')-\frac{M}{\beta},\\
\label{xp25h}
G_{p_xp_y}^{{\bf p}_0,{\bf x}_0}(\tau,\tau')&=&\meanrp{\tpx(\tau)\tpy(\tau')}=
iM^2\dg_{xx}^{{\bf x}_0}(\tau,\tau')-M^2\dgd_{xy}^{{\bf x}_0}(\tau,\tau')+\frac{1}{4}
M^2\osa^2G_{xy}^{{\bf x}_0}(\tau,\tau'),\\
\label{xp25i}
G_{p_zp_z}^{{\bf p}_0,{\bf x}_0}(\tau,\tau')&=&\meanrp{\tpz(\tau)\tpz(\tau')}=
-M^2\tdgd_{zz}^{{\bf x}_0}(\tau,\tau')-\frac{M}{\beta},
\end{eqnarray}
where the expectation values are defined by Eq.~(\ref{vpt07}). Note that all these 
Green functions are invariant under time translations such that
\begin{equation}
\label{xp26}
G_{\mu\nu}^{{\bf p}_0,{\bf x}_0}(\tau,\tau')=G_{\mu\nu}^{{\bf p}_0,{\bf x}_0}(\tau-\tau')
\end{equation}
with $\mu,\nu\in \{x,y,z,p_x,p_y,p_z\}$.

It is quite instructive to prove that all these Green functions can be decomposed into
a quantum statistical and a classical part as we did it in Eq.~(\ref{app19}). Since we
know that the classical correlation functions do not depend on the euclidean time,
all derivative terms in Eqs.~(\ref{xp25a})--(\ref{xp25i}) do not contribute a classical
term. We can write each Green function
\begin{equation}
\label{xp27}
G_{\mu\nu}^{{\bf p}_0,{\bf x}_0}(\tau,\tau')=G_{\mu\nu}^{\rm qm}(\tau,\tau')-
G_{\mu\nu}^{\rm cl},
\end{equation}
This relation has been already checked for 
Eqs.~(\ref{xp25a})-(\ref{xp25c}) in Appendix~\ref{appgen}. The classical contribution
is zero in Eqs.~(\ref{xp25d}), (\ref{xp25f}), and (\ref{xp25h}) following from the
absence of classical terms in derivatives of the Green functions and mixed
correlations like (\ref{app22a}). It seems surprising that the correlation (\ref{xp25e})
contains a classical term while (\ref{xp25d}) possesses none. This is, however, a consequence
of the cross product of the orbital angular momentum appearing in the action~(\ref{xp02})
and the explicit classical calculation entails
\begin{equation}
\label{xp28}
G_{xp_x}^{\rm cl}=\langle xp_x\rangle^{\rm cl}=0,
\qquad G_{xp_y}^{\rm cl}=\langle xp_y\rangle^{\rm cl}=\frac{2}{\beta}\,\frac{\osa}{\osb^2-\osa^2},
\end{equation}
where the latter is the subtracted classical term in Eq.~(\ref{app19}) when considering 
the first two substitutions in (\ref{xp13}). In Eq.~(\ref{xp25i}), the second term
is obviously the classical one since 
\begin{equation}
\label{xp29}
G_{p_zp_z}^{\rm cl}=\langle p_zp_z\rangle^{\rm cl}=\frac{M}{\beta}.
\end{equation}
The extraction of the classical terms 
\begin{equation}
\label{xp30}
G_{p_xp_x}^{\rm cl}=\langle p_xp_x\rangle^{\rm cl}=\frac{M}{\beta}\left(1+
\frac{\osa^2}{\osb^2-\osa^2}\right)
\end{equation}
in the case of the Green function $G_{p_xp_x}^{{\bf p}_0,{\bf x}_0}(\tau,\tau')$ 
requires the consideration of the last two terms in Eq.~(\ref{xp25g}). Thus we have shown that
the decomposition (\ref{xp27}) holds for each of the Green functions 
(\ref{xp25a})--(\ref{xp25i}). Note the necessity of subtracting the classical terms since
they all diverge in the classical limit of high temperatures ($\beta\to 0$). 
\end{appendix}

\end{document}